\documentstyle[aps,preprint,tighten,psfig]{revtex}
\begin{document}
\draft
\preprint{\vbox{\it Submitted to Nucl. Phys. A \hfill\rm TRI-PP-95-84}}

\title{Quasifree pion electroproduction from nuclei in the $\Delta$ region}
\author{Frank X. Lee}
\address{TRIUMF, 4004 Wesbrook Mall,
Vancouver, British Columbia, Canada V6T 2A3}
\author{Louis E. Wright}
\address{Institute of Nuclear and Particle Physics,
Department of Physics, Ohio University, \\ Athens, Ohio 45701, USA}
\author{C. Bennhold}
\address{Center for Nuclear Studies, Department of Physics,
The George Washington University, Washington, DC 20052, USA}
\date{\today}
\maketitle

\begin{abstract}
We present calculations of the reaction $A(e,e^\prime \pi N)B$
in the distorted wave impulse approximation.
The reaction allows for the study of the production process
in the nuclear medium
without being obscured by the details of nuclear transition densities.
First, a pion electroproduction operator suitable for nuclear
calculations is obtained by extending the Blomqvist-Laget
photoproduction operator to the virtual photon case.
The operator is gauge invariant, unitary,
reference frame independent,
and describes the existing data reasonably well.
Then it is applied in nuclei to predict nuclear cross sections
under a variety of kinematic arrangements.
Issues such as the effects of gauge-fixing,
the interference of the $\Delta$ resonance with the background,
sensitivities to the quadrupole component of the $\Delta$ excitation
and to the electromagnetic form factors,
the role of final-state interactions, are studied in detail.
Methods on how to experimentally separate
the various pieces in the coincidence cross section are suggested.
Finally, the model is compared to a recent SLAC experiment.
\end{abstract}
\vspace{1cm}

\parskip=2mm
\section{Introduction}
\label{intro}

The goal of this work is to develop a theoretical framework for
analyzing the exclusive, quasifree pion electroproduction
from complex nuclei, denoted by $A(e,e^\prime \pi N)B$,
in the $\Delta$(1232) resonance region.
Here, ``exclusive'' means that the outgoing particles
$e^\prime$, $\pi$, $N$ are detected
in coincidence, and the nucleus undergoes transitions to discrete
final nuclear states, usually the ground state.
The term ``quasifree'' refers to those processes
which can be identified as
taking place on a single nucleon inside the nucleus
(impulse approximation).

\begin{figure}
\centerline{\psfig{file=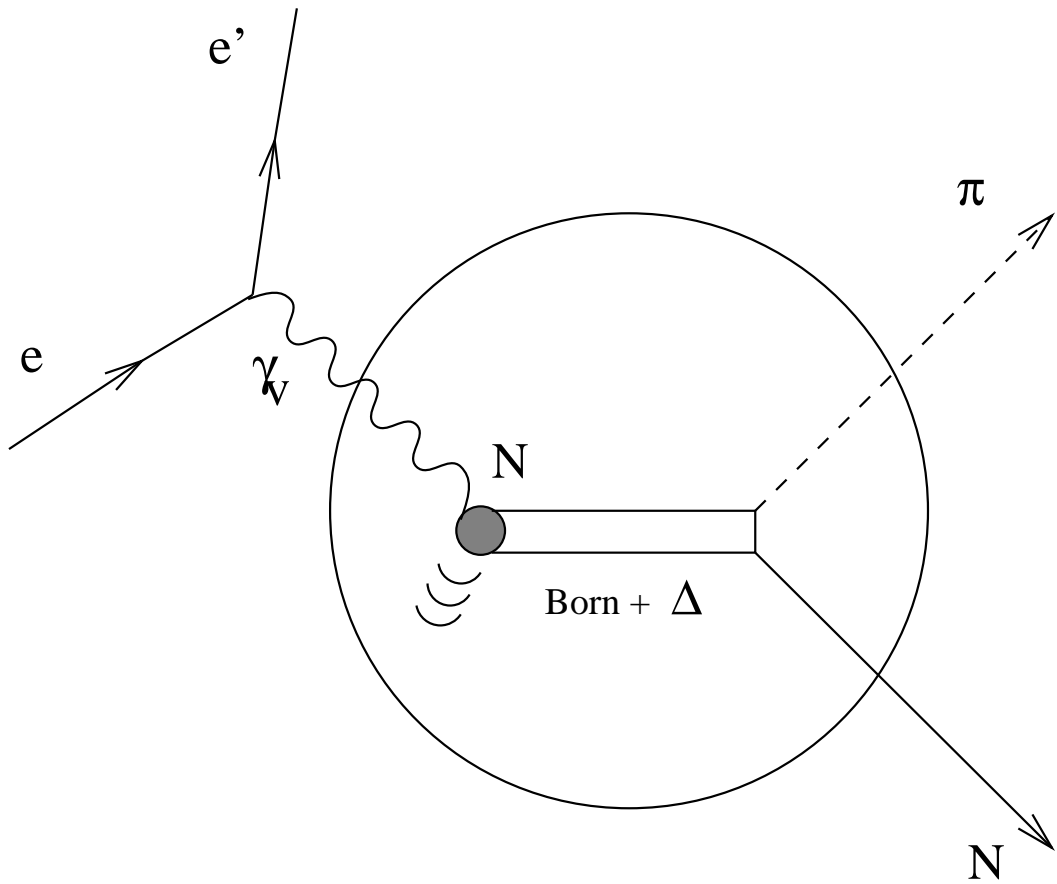,width=14cm}}
\vspace{1cm}
\caption{An illustration of the reaction
$A(e,e^\prime \pi N)B$ in the $\Delta$ region.
\label{eepic}}
\end{figure}

The study of this reaction is of interest because
it provides a testing ground for our knowledge on
several areas of nuclear physics research:
the electromagnetic production of pions from nucleons;
quasi-elastic electron scattering off nuclei;
the pion-nucleus and the nucleon-nucleus interaction.
The combination is best appreciated by visualizing the
photoproduction process in the impulse approximation
with a single-particle model of the nucleus.
In this picture (see Fig.~\ref{eepic}),
the incident photon (real or virtual) penetrates
the nucleus and couples to an individual nucleon via the latter's charge
and magnetic moment. This causes the nucleon to oscillate and radiate
pions. The produced pions, along with the nucleons
that also exit the system,
subsequently rescatter from the remaining
nucleons before finally escaping and reaching the detector.
The initial pion production stage involves knowledge of the
photoproduction of pions off single nucleons,
the rescattering stage requires an understanding
of the way pions and nucleons scatter off nuclei,
while the overall transition of the nucleus between specific initial and
final states is the subject of electron scattering studies below
pion threshold.
In principle, each of these ingredients can become the subject
of theoretical scrutiny by using the best knowledge on the
other ingredients.
The most important motivation, however, is to use this reaction to
study the $\Delta$ excitation and propagation in the nuclear medium.

Traditionally, pion photoproduction from nuclei has been studied
with the reaction $A(\gamma,\pi)B$~\cite{book}.
Since this reaction requires the
final nucleon to remain bound in the residual nucleus,
it involves relatively high momentum transfers.
Consequently, these processes are very sensitive to the
details of the nuclear transition densities which
often obscure the process of primary interest: pion photoproduction
in the medium. In recent years, it is recognized that sensitivities
to the nuclear transition matrix elements can be greatly reduced
by allowing the final nucleon involved in the
pion production process to exit the nucleus, namely,
by studying the coincidence reaction $A(\gamma, \pi N)B$.
This is mainly due to the quasifree nature of the reaction:
the momentum transfer to the residual nucleus can be made small.
By measuring in coincidence the decay products of the $\Delta$,
one can get a much better handle on the $\Delta$ in the medium.
Comparison with existing data already showed good promise of
such reactions~\cite{bates,li,sato}.

Pion electroproduction from nuclei in the $\Delta$ region,
on the other hand, has been studied in less detail.
While the reaction $A(e,e^\prime \pi)B$ has been investigated
in Ref.~\cite{st}, to our knowledge
there is no theoretical work on the process $A(e,e^\prime \pi N)B$ in
the $\Delta$ region.
The use of electrons instead of photons
adds the advantage of probing the longitudinal response and
longitudinal-transverse interference, but with the price of
dealing with more complicated structure in the cross section.

Such coincidence experiments are challenging
because the final nuclear state, the outgoing pion and
nucleon all need to be identified in coincidence with
sufficient energy resolution and solid angle coverage.
Data for $A(\gamma,\pi N)B$ in the $\Delta$ region are sparse~\cite{steen}.
Recently, two experiments on the reaction have just been
completed and data are being analyzed~\cite{nik,legs}.
Data on $A(e,e^\prime \pi N)B$ are almost non-existent,
until the first experiment was recently carried out at SLAC~\cite{slac}.
However, with the advent of new
high duty cycle accelerators such as LEGS at Brookhaven,
Bates, NIKHEF, MAMI and CEBAF,
the situation is expected to greatly improve in the near future.

In section~\ref{electro}, we describe our pion electroproduction
operator and show how it compares with existing data.
In section~\ref{dwia},
the DWIA formalism for the reaction $A(e,e^\prime \pi N)B$ is
derived.
In section~\ref{result}, we show results under a variety of kinematic
arrangements in order to expose different aspects of the physics
involved, and compare with the SLAC experiment.
Section~\ref{sumcon} gives a summary of our findings.
The full operator is given in Appendix~\ref{blvirtual}.
The electromagnetic form factors used in the calculation are
given in Appendix~\ref{formfactors}.

\section{Pion electroproduction on the nucleon}
\label{electro}

In order to study pion electroproduction in the resonance
region on complex nuclei,
one first has to understand the production process on the nucleon.
To this end there have been extensive studies,
for example, in~\cite{cgln,den,ber,shaw,book1,mw,jl,chris,gar,nl,phd}.
The expressions for the elementary cross sections have become standard.
We have rederived them as a by product in our derivation of
the DWIA formalism for the nuclear case~\cite{phd} using the
density matrix method.
This also serves as a consistency check.
In the following, we will skip the derivation and only
give the expressions relevant for our discussions.

\subsection{Differential cross section}

All quantities are in the laboratory frame unless
otherwise mentioned
(c.m. quantities are explicitly denoted by a superscript c).
Due to the relative weakness of the electromagnetic interaction,
electron scattering can be treated as the exchange of a virtual
photon which carries energy $\omega=E_e-E_{e^\prime}$ and momentum
${\bf k}={\bf p}_e -{\bf p}_{e^\prime}$. The 4-momentum transfer squared
is spacelike and given by
\begin{equation}
k^2 =\omega^2-|{\bf k}|^2=
 -4E_e E_{e^{\prime}}\sin^2\frac{\theta_e}{2}
\label{k2}
\end{equation}
where $\theta_e$ is the electron scattering angle and the
electron mass is neglected.
The differential cross section
for unpolarized pion electroproduction on single nucleon can be written
as an electron flux factor
times the virtual photon cross section :
\begin{equation}
\frac{d^3\sigma}{d\Omega_{e^{\prime}}dE_{e^{\prime}}d\Omega_{\pi}}
= \Gamma \frac{d\sigma_v}{d\Omega_{\pi}}
\label{free0}
\end{equation}
where
\begin{equation}
\frac{d\sigma_v}{d\Omega_{\pi}}=
\frac{d\sigma_{\scriptscriptstyle T}}{d\Omega_{\pi}}
+ \epsilon\,\frac{d\sigma_{\scriptscriptstyle L}}{d\Omega_{\pi}}
+ \epsilon\,\frac{d\sigma_{\scriptscriptstyle TT}}{d\Omega_{\pi}}
+\sqrt{\epsilon(1+\epsilon)}\,
\frac{d\sigma_{\scriptscriptstyle TL}}{d\Omega_{\pi}}.
\label{free}
\end{equation}
The flux factor is defined by
\begin{equation}
\Gamma= \frac{\alpha}{2{\pi}^2}\,\frac{E_{e^{\prime}}}{E_e}\,
\frac{K}{(-k^2)}\,\frac{1}{1-\epsilon}
\end{equation}
In this expression, $\alpha$ is the fine structure constant,
$K =(W^2 -m^2)/{2m}$, where $m$ is the nucleon mass, is called
the virtual photon equivalent energy and $W^2=k^2+m^2+2m\omega$
is the invariant mass squared of the $\pi$N pair.
The variable $\epsilon$ is the degree of transverse
polarization of the virtual photon and is given by
\begin{equation}
\epsilon=\left(1+\frac{2|{\bf k}|^2}{-k^2}\,
\tan^2 \frac{\theta_e}{2} \right)^{-1}.
\label{eps}
\end{equation}
Note that $\epsilon$ is invariant with respect to the Lorentz boost
along the virtual photon direction ${\bf k}$.
The transverse, longitudinal, polarization and interference
cross sections, as they are called,
 are given in terms of the hadronic current matrix elements by
\begin{equation}
\frac{d\sigma_{\scriptscriptstyle T}}{d\Omega_{\pi}}
=\frac{A}{2} \sum_{m_i m_f} \left(|\langle J_x\rangle |^2 +|\langle J_y\rangle
|^2 \right),
\label{t}
\end{equation}
\begin{equation}
\frac{d\sigma_{\scriptscriptstyle L}}{d\Omega_{\pi}}=
A\,\frac{-k^2}{\omega^2}\,\sum_{m_i m_f}\,|\langle J_z\rangle |^2,  \label{l}
\end{equation}
\begin{equation}
\frac{d\sigma_{\scriptscriptstyle TT}}{d\Omega_{\pi}}
=\frac{A}{2} \sum_{m_i m_f} \left(|\langle J_x\rangle |^2 -|\langle J_y\rangle
|^2 \right),
\end{equation}
\begin{equation}
\frac{d\sigma_{\scriptscriptstyle TL}}{d\Omega_{\pi}}=\frac{-A}{2}
\sqrt{\frac{-2k^2}{\omega^2}}\,
\sum_{m_i m_f} \left( \langle J_x\rangle ^{*}\langle J_z\rangle
+\langle J_x\rangle  \langle J_z\rangle ^{*} \right)
\label{i}
\end{equation}
where the z-axis is along ${\bf k}$.
Here the notation $\langle J_\mu\rangle  \equiv \langle
m_f|\,J_\mu\,|m_i\rangle $ represents the
matrix elements between initial and final nucleon states.
Eq.~(\ref{t}) to Eq.~(\ref{i}) are valid both in the laboratory frame
and in the c.m. frame.
The kinematic factor A is given in the laboratory frame by
\begin{equation}
A=\frac{m|{\bf q}|}{32{\pi}^2K|E_N+E_\pi({\bf q}-{\bf k})
\cdot{\bf q}/|{\bf q}|^2|},
\end{equation}
and in the c.m. frame by
\begin{equation}
A=\frac{|{\bf k}|}{2K}\frac{|{\bf q}^c|}{|{\bf k}^c|}
\left(\frac{m}{4\pi W}\right)^2
\end{equation}
where the pion four momentum $q^\mu=(E_\pi,{\bf q})$.
The above definitions ensure that in the real photon limit
$k^2 \rightarrow 0$, the transverse cross section Eq.~(\ref{t})
goes to the real photon cross section.

Two comments are in order.
First,  a different form
is sometimes used in the literature where the explicit pion
angular dependence is written out:
\begin{equation}
\frac{d\sigma_v}{d\Omega_{\pi}} =
\frac{d\sigma_{\scriptscriptstyle T}}{d\Omega_{\pi}}
+ \epsilon\,\frac{d\sigma_{\scriptscriptstyle L}}{d\Omega_{\pi}}
+ \epsilon\,\frac{d\sigma_{\scriptscriptstyle TT}}{d\Omega_{\pi}}
\,\sin^2{\theta_\pi}\,\cos(2\phi_\pi)
+\sqrt{\epsilon(1+\epsilon)}\,
\frac{d\sigma_{\scriptscriptstyle TL}}{d\Omega_{\pi}}
\,\sin{\theta_\pi}\,\cos{\phi_\pi}.
\label{explicit}
\end{equation}
Under this definition the polarization and interference cross sections will
differ from those in Eq.~(\ref{free}) by the extra angular factors.
Therefore one has to be careful about the definition used when
comparing with data or with other calculations.
It is obvious from this expression that the total virtual cross section
has contributions only from the transverse and longitudinal terms:
\begin{equation}
\sigma_v=\sigma_{\scriptscriptstyle T}+\epsilon\,
\sigma_{\scriptscriptstyle L}.
\label{total}
\end{equation}

Second, we discuss a kinematic singularity problem in the c.m. frame.
The virtual photon energy in the c.m. frame
can be written as
$\omega^c=(W^2+k^2-m^2)/(2W)$. Since the virtual photon is
spacelike, $\omega^c$ can vanish or even become negative.
This would lead to apparent singularities in the cross sections
(see Eqs.~(\ref{l}) and (\ref{i})).
For pion electroproduction, $W \geq m+m_\pi \sim$ 1080 MeV,
so it could happen for 4-momentum transfer starting at
$-k^2 = m^2_\pi+2m m_\pi \sim 0.24\; (GeV/c)^2$ which lies mostly
in the region of interest. In the $\Delta$ region, for example,
W=1230 MeV, then $\omega^c$ vanishes at approximately
$-k^2=0.63\; (GeV/c)^2$.
One obvious way to circumvent the problem is to work
in the laboratory frame.
However, since all the electroproduction data are conventionally presented
in the c.m. frame, we still need to transform the cross sections
into the c.m. frame. This can be achieved by the following Jacobian:
\begin{equation}
\frac{d\Omega_{\pi}}{d\Omega^c_{\pi}}
=\frac{\gamma |{\bf q}|^c}{|{\bf q}|}
(1-\frac{E_\pi}{|{\bf q}|}\beta \cos{\theta_\pi})
\end{equation}
where the Lorentz boost from the laboratory frame to the
c.m. frame is $\mbox{\boldmath $\beta$}={\bf k}/(\omega +m)$.
We will give another solution below when we
discuss gauge invariance.

\subsection{The electroproduction operator}

Traditionally, models of pion electroproduction on the nucleon are built
in terms of CGLN amplitudes~\cite{cgln,den,ber}
by means of a multipole decomposition.
These amplitudes are expressed in the c.m. frame.
A transformation is needed in order to use them in nuclear calculations
where the struck nucleon has a momentum distribution (Fermi motion).
Such a transformation is complicated and sometimes ambiguous.
Moreover, it is not clear how to extrapolate the amplitudes to off-shell.

Our goal is to have an operator that can describe the
elementary process reasonably well and that is suitable for
application to nuclear calculations.
To this end, we employed the techniques of Blomqvist
and Laget (BL)~\cite{bl1,laget} to construct an electroproduction
operator. The BL photoproduction operator is based on
an effective Lagrangian approach, incorporating gauge invariance and
unitarity. It is expressed in an arbitrary frame of reference
which makes it convenient for use in nuclear calculations.
The diagrammatic structure of the amplitude provides a
physically transparent picture of the individual processes which
contribute and facilitates the discussion of nonlocality
and off-shell effects.
The BL amplitudes describe the elementary pion photoproduction
data reasonably well over a wide range of energies
and have enjoyed success in many nuclear calculations~\cite{book}.
These features make it a good candidate for our purposes.
In recent years, attempts to upgrade the BL model have been
made~\cite{dmw} and models based on the Hamiltonian approach
have been developed~\cite{nbl}.

The procedure of extending the BL pion photoproduction operator
to electroproduction is straightforward.
Firstly, the same Feynman diagrams
as in photoproduction are considered,
but electromagnetic form factors describing the charge
and magnetic moment distribution of
the particles need to be introduced at appropriate
vertices to replace the static charge and magnetic moments.
They are the pion form factor $F_\pi(k^2)$, the nucleon form
factors $F_1(k^2)$ and $F_2(k^2)$ (or equivalently $G_M(k^2)$ and
$G_E(k^2)$), the nucleon axial form factor $F_A(k^2)$, the
$\Delta$ resonance form factor $F_\Delta (k^2)$, and in the case of
$\omega$ meson exchange, the $\omega$ form factor $F_\omega (k^2)$.
Secondly, terms proportional to $\epsilon \cdot {\bf k}$ that were dropped
in photoproduction need to be included
since they are non-vanishing due to the additional longitudinal
polarization of the virtual photon.
Thirdly, in addition to the transverse quadrupole E2 term,
there is the longitudinal quadrupole C2 term (sometimes called L2).
To improve the readability of the paper,
the full expressions for the obtained operator are given
in Appendix~\ref{blvirtual} and the form factors used in the model are
given in Appendix~\ref{formfactors}.

\subsection{Gauge invariance}

When electromagnetic form factors of the various particles are introduced
at each vertex, the Born terms in the resulting interaction are no longer
 gauge invariant (the delta term is constructed
to be separately gauge invariant and hence not affected).
Current conservation
\begin{equation}
J_0=|{\bf k}|\, J_z/\omega
\end{equation}
has been assumed when deriving the formalism in the
previous section.
As the magnitude of $k^2$
increases the form factors become quite different from each other,
especially the pion form factor.
Thus, restoration of gauge invariance is needed when
different form factors are used in the calculation.  While
various solutions to this problem have been explored, we use
a method of restoring gauge invariance which
does not change the interaction in the Lorentz gauge.

  Let us assume we have written the interaction in the Lorentz gauge
as $\varepsilon_\mu J^\mu$.  We can check gauge invariance
(or more precisely current conservation)
by replacing  $\varepsilon_\mu$ by the photon four
 momentum $k_\mu$ and checking if $k_\mu J^\mu =0$.
If $J^\mu$ contains general form factors then this equation will not be
satisfied.  However, current conservation can be restored by
 defining a new current
\begin{equation}
{J^{\prime}}^{\mu}=J^{\mu}-k^{\mu}(k_{\nu}J^{\nu})/{k^2}.
\label{gauge}
\end{equation}
Furthermore, in the Lorentz gauge
$\varepsilon_\mu {J^{\prime}}^\mu =\varepsilon_\mu J^\mu$ since
the Lorentz gauge condition is $\varepsilon_\mu k^\mu=0$.   Restoring
current conservation by this method remains {\em ad hoc}, but it seems
to do minimal damage, and allows us to use realistic form factors. With
a conserved current we have regained gauge invariance and thus can carry
out the calculation in the most convenient gauge.  Once better quality
experimental data is available, the validity of this approach can be
investigated. For the present we will begin with Laget's
non-relativistic parameterization of pion photoproduction
with gauge fixing to the same order in $(p/m)$
and check if our different form factors and additional current terms
continue to fit the available data.
The resulting gauge fixing terms in Eq.~(\ref{gauge})
are given in Appendix~\ref{blvirtual}.

As discussed above,
the vanishing virtual photon energy in the c.m. frame
is problematic since the transformation to eliminate the time
component of the hadronic current by current conservation
becomes singular at this point.
It is not a true singularity in the sense that one can
in principle factorize out $\omega$ in $J_z$ to explicitly cancel
the $\omega$ factor in front. But in practice that is cumbersome
and sometimes not possible.

An alternative is to keep all four components of the current.
As a result~\cite{phd}, the transverse cross section
$d\sigma_{\scriptscriptstyle T}/d\Omega_{\pi}$ and
polarization cross section
$d\sigma_{\scriptscriptstyle TT}/d\Omega_{\pi}$
remain unchanged, while the longitudinal and interference cross sections
take on new forms:
\begin{equation}
\frac{d\sigma_{\scriptscriptstyle L}}{d\Omega_{\pi}}
= A\,\sum_{m_i m_f} \left[
\frac{|{\bf k}|^2}{-k^2}\,|\langle J_0\rangle |^2
+\frac{\omega^2}{-k^2}\,|\langle J_z\rangle |^2
-\frac{|{\bf k}|\omega}{-k^2}\,\left(\langle J_z\rangle ^{*}\langle J_0\rangle
+\langle J_0\rangle ^{*}\langle J_z\rangle  \right) \right],
\label{l44}
\end{equation}
and
\begin{equation}
\frac{d\sigma_{\scriptscriptstyle TL}}{d\Omega_{\pi}} =
\frac{A}{2} \sum_{m_i m_f} \left[
\frac{\omega}{\sqrt{-2k^2}}\,\left(\langle J_z\rangle ^{*}\langle J_x\rangle
+\langle J_x\rangle ^{*}\langle J_z\rangle \right)
-\frac{|{\bf k}|}{\sqrt{-2k^2}}\,\left(\langle J_0\rangle ^{*}\langle
J_x\rangle
+\langle J_x\rangle ^{*}\langle J_0\rangle  \right) \right].
\label{i44}
\end{equation}
Clearly, there is no singularity in this formulation,
no matter what reference frame is used.
Since no current conservation is assumed in the derivation,
one can work in any gauge that is convenient.
One can also use the new equations to
investigate the degree of gauge invariance violation before gauge
fixing.
On the other hand, if one uses current conservation
to eliminate $J_0$, Eqs.~(\ref{l44}) and~(\ref{i44})
will reduce to Eq.~(\ref{l}) and Eq.~(\ref{i}), respectively,
as they should.

\subsection{Comparison with data}

In this section we compare theoretical calculations on the nucleon
performed with the electroproduction operator
with a large body of existing
data~\cite{batzner,fischer,bardin,breuker,bonn},
most of which were obtained in the 1970s at Bonn and Saclay.
Since our aim is to use the operator in nuclei,
we only select some representative figures
to show the overall quality of the operator.
More details can be found in Ref.~\cite{phd}.

\begin{figure}
\centerline{\psfig{file=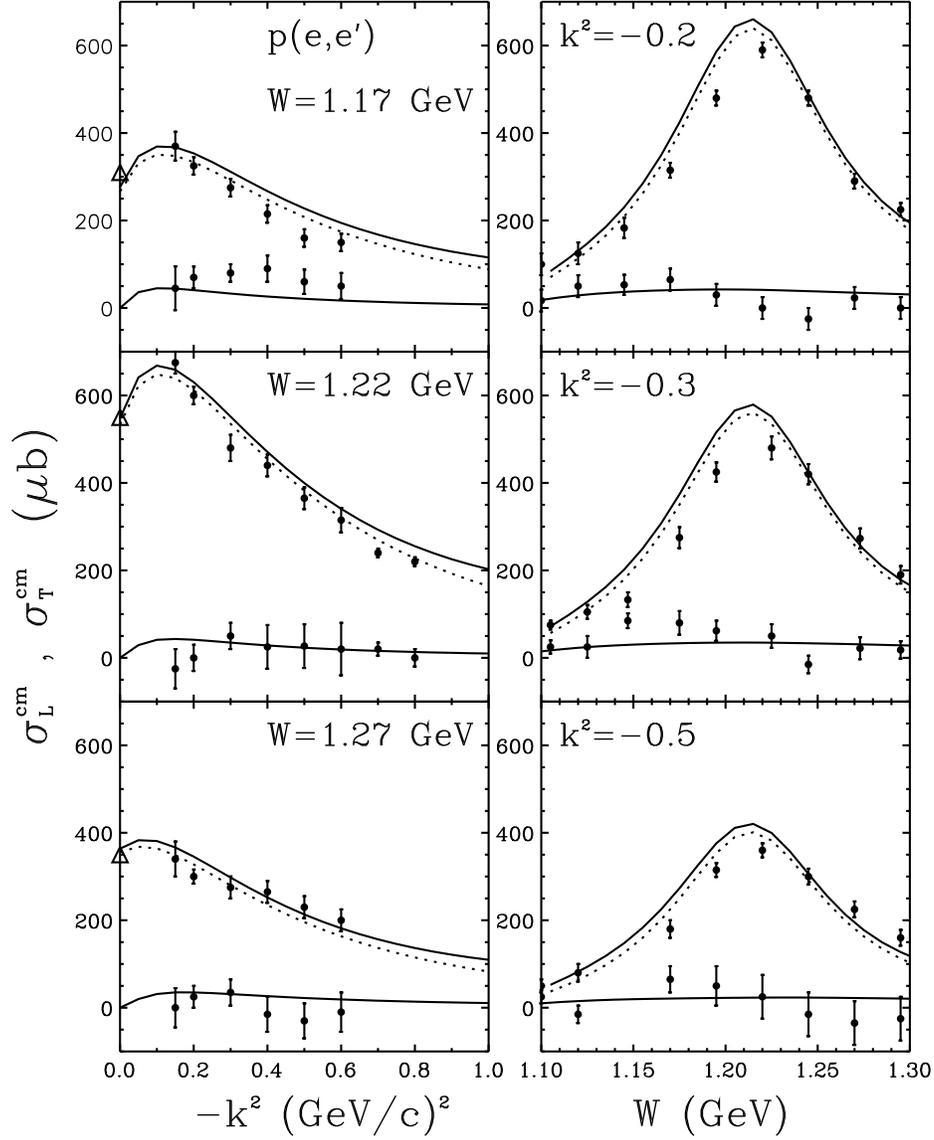,width=14cm}}
\vspace{1cm}
\caption{The $k^2$ and W dependence of the separated
total cross sections for the
inclusive reaction $p(e,e^\prime)$ is compared to the
data~\protect\cite{batzner}.
The upper curves are for the transverse cross section $\sigma_T$,
lower curves for longitudinal cross section $\sigma_L$. The solid curves are
calculated in the laboratory frame, whereas the dotted curves in the
c.m. frame. The triangles are the real photon
points~\protect\cite{fischer}.}
\label{tot}
\end{figure}

Fig.~\ref{tot} shows the comparison with the total cross section data
 on $p(e,e^\prime)$ in the $\Delta$ region.
This experiment was inclusive---only electrons were detected.
Thus, the curves represent the combined contributions from
both the $ep \rightarrow n \pi^+$ and the $ep \rightarrow p \pi^0$ processes.
Since the total virtual cross section only consists of longitudinal and
transverse contributions (see Eq.~(\ref{virtual})),
a separation of them is possible by varying electron kinematics.
Several interesting features can be seen in the figure.
First, the $\Delta$ resonance peaks are clearly present in the W dependence.
Second, the longitudinal cross sections are consistently small
compared to the transverse cross section.
Third, in the real photon limit $k^2 \rightarrow 0$, the calculations
agree with the measured real photon points.
Fourth, the operator is supposed to be valid in any reference frame,
which is indeed the case judging by the small differences between
the solid lines and the dotted lines.
The overall agreement with the total cross section data is satisfactory.

\begin{figure}
\centerline{\psfig{file=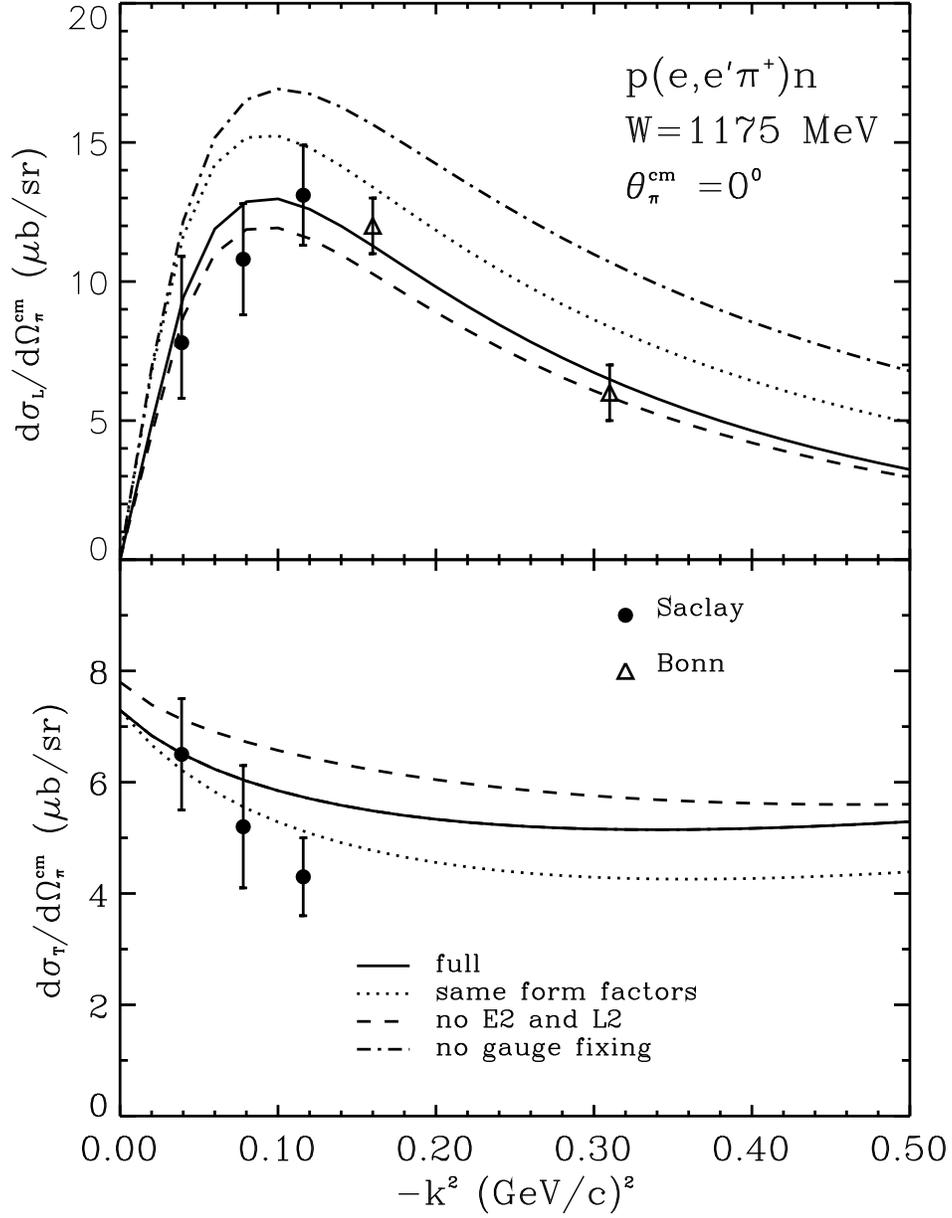,width=14cm}}
\vspace{1cm}
\caption{The separated longitudinal and transverse
differential cross sections
of the reaction $p(e,e^\prime \pi^+)n$ are plotted as function of the virtual
photon four-momentum squared $k^2$ at $\theta_\pi=0$ and W=1175 MeV.
The data points are from Saclay~\protect\cite{bardin}
and Bonn~\protect\cite{breuker}.}
\label{lt}
\end{figure}

Fig.~\ref{lt} shows the separated longitudinal and transverse cross
sections with pions exiting along the direction of the virtual photon
momentum. In such kinematics the polarization and the interference cross
sections vanish (see Eq.~(\ref{explicit})), allowing the
longitudinal-transverse separation by varying $\epsilon$.
The solid curves represent the calculation with the full operator which
uses realistic form factors, contains the E2 and L2 terms and includes the
gauge fixing terms. The dotted curves are calculated with the same
form factors $F_\pi=F_A=F_1$, in which case the Born terms are already
gauge invariant, thus there is no need to restore gauge invariance.
 The dashed curves are the same as for the full operator,
except the quadrupole terms E2 and L2 have been omitted.
The dash-dotted curves show the effect of leaving out
the gauge fixing terms. Clearly, gauge
restoration which only affects the longitudinal term
is important in improving the agreement between the model
and the data.

The remaining figures in this section compare three calculations:
the full operator (solid lines), no gauge fixing terms
(dotted lines), no E2 and L2 terms (dashed line), in order to
study the significance of
gauge fixing and the quadrupole excitation of the $\Delta$.
The definition in Eq.~(\ref{explicit}) is used for
the polarization cross section
$d\sigma_{\scriptscriptstyle TT}/d\Omega_{\pi}$ and
the interference cross section
$d\sigma_{\scriptscriptstyle TL}/d\Omega_{\pi}$ in these figures.

\begin{figure}
\centerline{\psfig{file=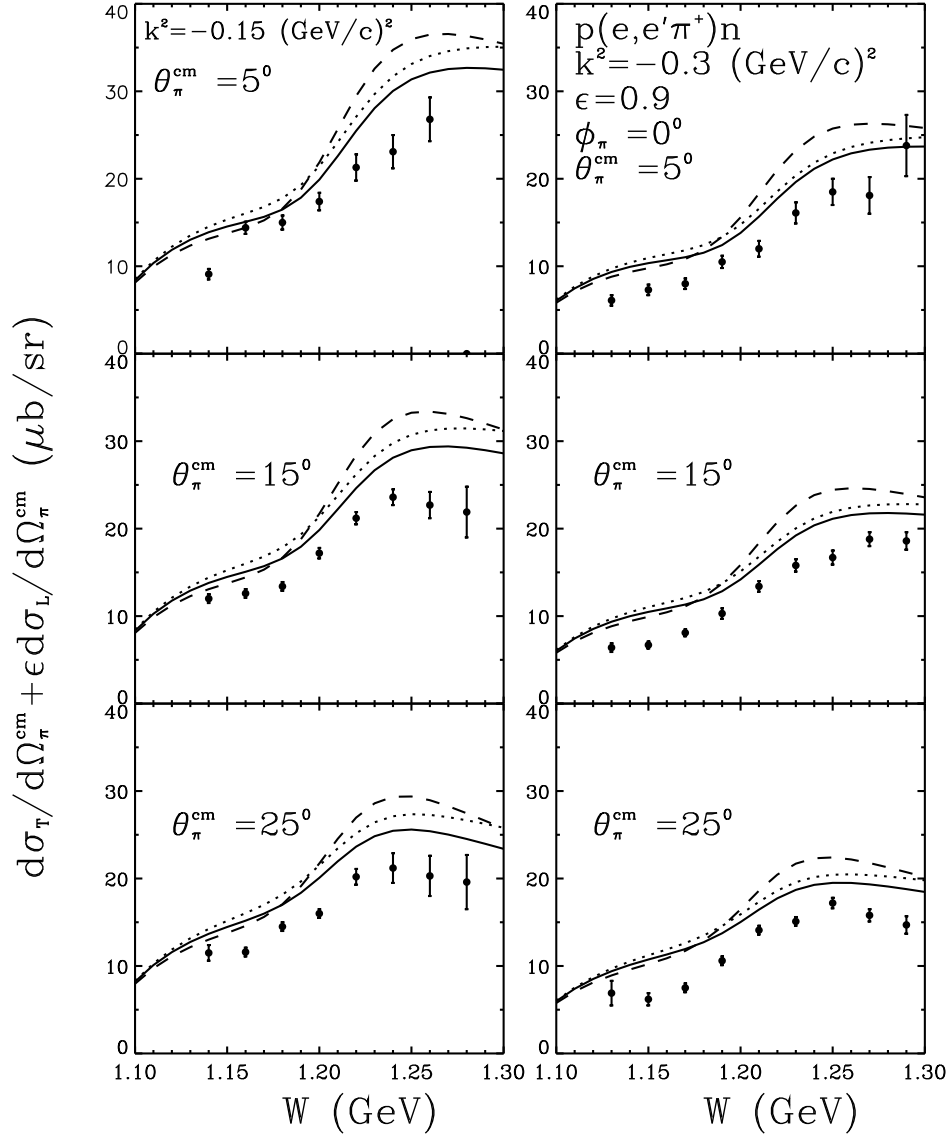,width=14cm}}
\vspace{1cm}
\caption{Separated cross sections $d\sigma_{\scriptscriptstyle T}
/d\Omega_\pi+\epsilon\,d\sigma_{\scriptscriptstyle L}/d\Omega_\pi$ as a
function of W at fixed $k^2$ and $\theta_\pi$.
The solid lines are calculated with the full operator; dotted lines
with the same form factors and no gauge fixing terms; dashed lines with  the
quadrupole component left out. Data are from Ref.~\protect\cite{breuker}.}
\label{k1530tl}
\end{figure}

\begin{figure}
\centerline{\psfig{file=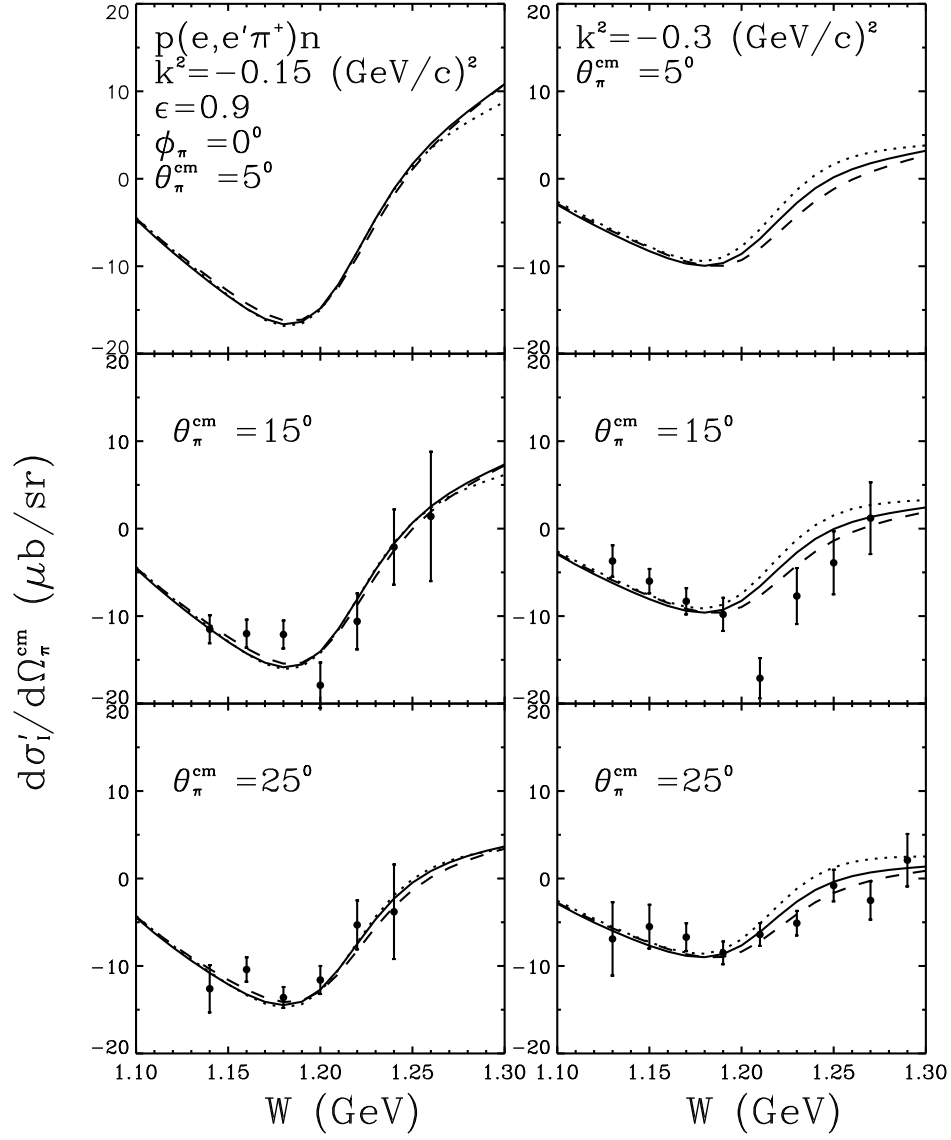,width=14cm}}
\vspace{1cm}
\caption{Same as in Fig.~\protect\ref{k1530tl},
but for the interference cross section
$d\sigma_{\scriptscriptstyle TL}/d\Omega_\pi$.}
\label{k1530i}
\end{figure}

Fig.~\ref{k1530tl} and Fig.~\ref{k1530i},
compares the W dependence of the
separated cross sections with data for different
values of $k^2$ and $\theta_\pi$.
Gauge fixing is not very significant in these cases,
although its inclusion improves the agreement between theory
and experiment.
There is some sensitivity to the E2 and L2 terms around W=1230 MeV,
right on top of the $\Delta$ resonance.

\begin{figure}
\centerline{\psfig{file=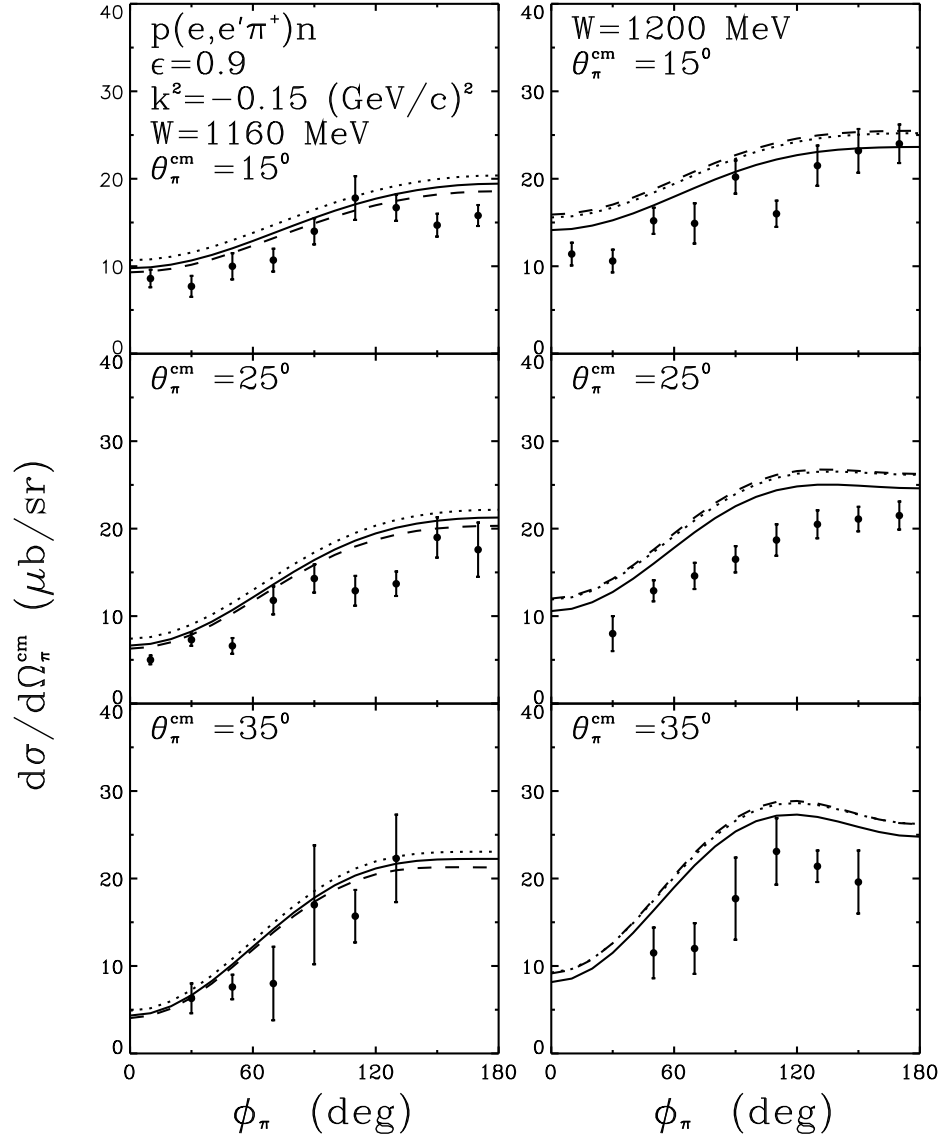,width=14cm}}
\vspace{1cm}
\caption{Pion azimuthal angular dependence of the
virtual photoproduction
cross sections are compared with the data~\protect\cite{breuker} at
$k^2=-0.15\;(GeV/c)^2$, W=1160, 1200 MeV
 and $\theta_\pi=15^0, \;25^0, \;35^0$.
The legends are the same as in Fig.~\protect\ref{k1530tl}.}
\label{ph11}
\end{figure}

In Fig.~\ref{ph11}, the pion azimuthal angle
distributions of the virtual cross sections (unseparated)
are presented along with data at different values of W, $k^2$, and
$\theta_\pi$.
Note that both gauge fixing and the quadrupole excitation exhibit more
significance in the $\Delta$ region.

\begin{figure}
\centerline{\psfig{file=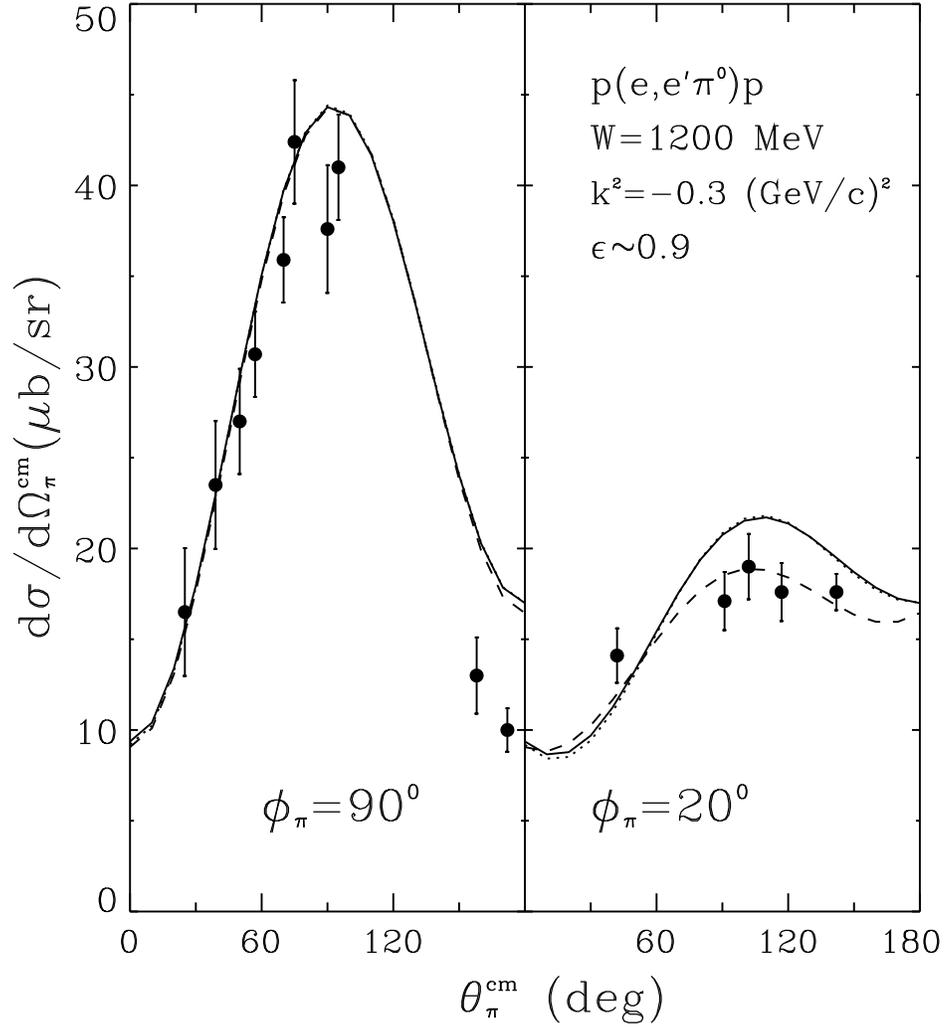,width=14cm}}
\vspace{1cm}
\caption{The $\theta_\pi$ distributions for the virtual cross section of
$p(e,e^\prime \pi^0)p$
are compared to the data from Bonn~\protect\cite{bonn}
for two pion azimuthal angles. The solid lines are calculated with the full
operator; dotted lines
with the same form factors and no gauge fixing terms; dashed lines without
the quadrupole component.}
\label{bonn}
\end{figure}

Fig.~\ref{bonn} shows the result of an analysis
of the $p(e,e^\prime \pi^0)p$ reaction performed at Bonn,
in a nearly coplanar ($\phi_\pi=20^0$) and a perpendicular
($\phi_\pi=90^0$) kinematic condition. The strong difference between these
two kinematics is due to the polarization cross section of which
the contribution changes sign when $\phi_\pi$ varies from 0 to $90^0$.
In perpendicular kinematics, the contribution of the
polarization cross section vanishes and the experiment determines the
transverse coupling of the $\Delta$. In nearly coplanar kinematics
the effects of the polarization cross section are at maximum and the
quadrupole component becomes more important.
Note, that the effects of gauge fixing in this case
 are small, almost indistinguishable from the full calculation.

\begin{figure}
\centerline{\psfig{file=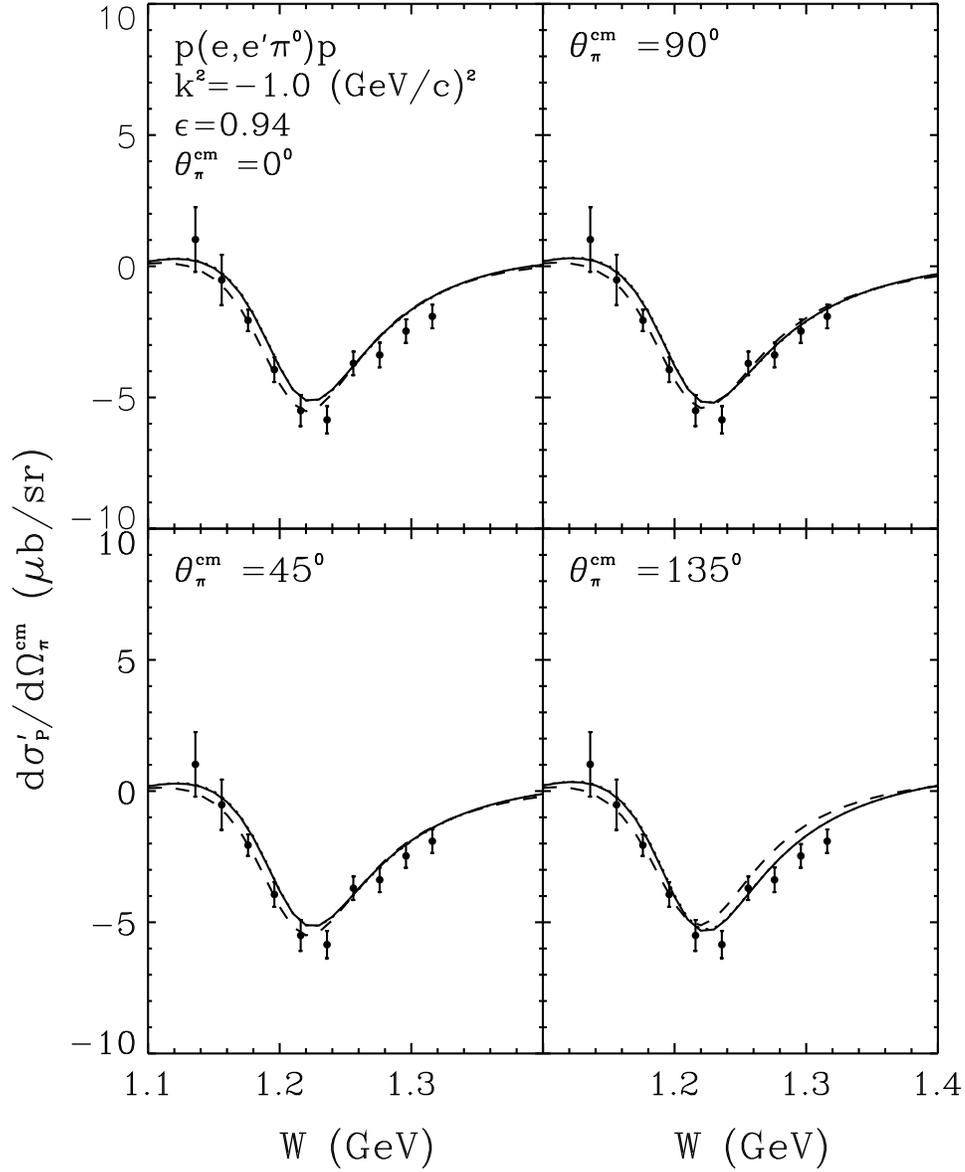,width=14cm}}
\vspace{1cm}
\caption{Polarization cross section
$d\sigma_{\scriptscriptstyle TT}/d\Omega_\pi$
as a function of W at $k^2=-1\;(GeV/c)^2$ and 4 values of $\theta_\pi$.
The solid lines are calculated with the full operator; dotted lines
with the same form factors and no gauge fixing terms; dashed lines without
 the quadrupole component. Data are from Ref.~\protect\cite{breuker}.}
\label{highp}
\end{figure}

Fig.~\ref{highp} presents the separated cross sections
at the even higher value of $k^2=-1\;(GeV/c)^2$.
The agreement with data in this case is excellent.

The overall agreement agreement between the model and the data
is satisfactory,
which is sufficient for our purposes in this exploratory study.
We stress that
we have not attempted to readjust any parameters in order to fit the data.
All the parameters are already fixed by the BL pion photoproduction operator.
In this sense the results are predictions rather than fits.
As new data become available in the near future, it may become
necessary to refine the operator by refitting, or to construct a
new operator suitable for use with a relativistic
treatment of the nucleons in the nucleus.

\section{The DWIA model for pion electroproduction on nuclei}
\label{dwia}

Having obtained the elementary transition
operator, we are now in the position to study
quasifree pion electroproduction on nuclei, $A(e,e^\prime \pi N)B$,
in the $\Delta$ region.
The term `quasifree' means that the process can be identified
as taking place on a single nucleon in nuclei.
This happens when the missing momentum (or the momentum transfer to
the target) is relatively small, say, less than $300\; MeV/c$.
In such a situation, the impulse approximation is expected to hold.
We employ the Distorted Wave Impulse Approximation (DWIA) which closely
follows our previous approach for photoproduction in Ref.~\cite{li}.
However, the new longitudinal polarization of the virtual photon
generates additional terms in the cross section which are nontrivial
in a full DWIA framework.
See Fig.~\ref{eepic} for an diagrammatic illustration of the reaction
$A(e,e^{\prime}\pi N)B$ in the $\Delta$ region in DWIA.
We first discuss the kinematics, then briefly describe the
nuclear structure inputs, and finally derive the cross sections.

\subsection{Kinematics}

\begin{figure}
\centerline{\psfig{file=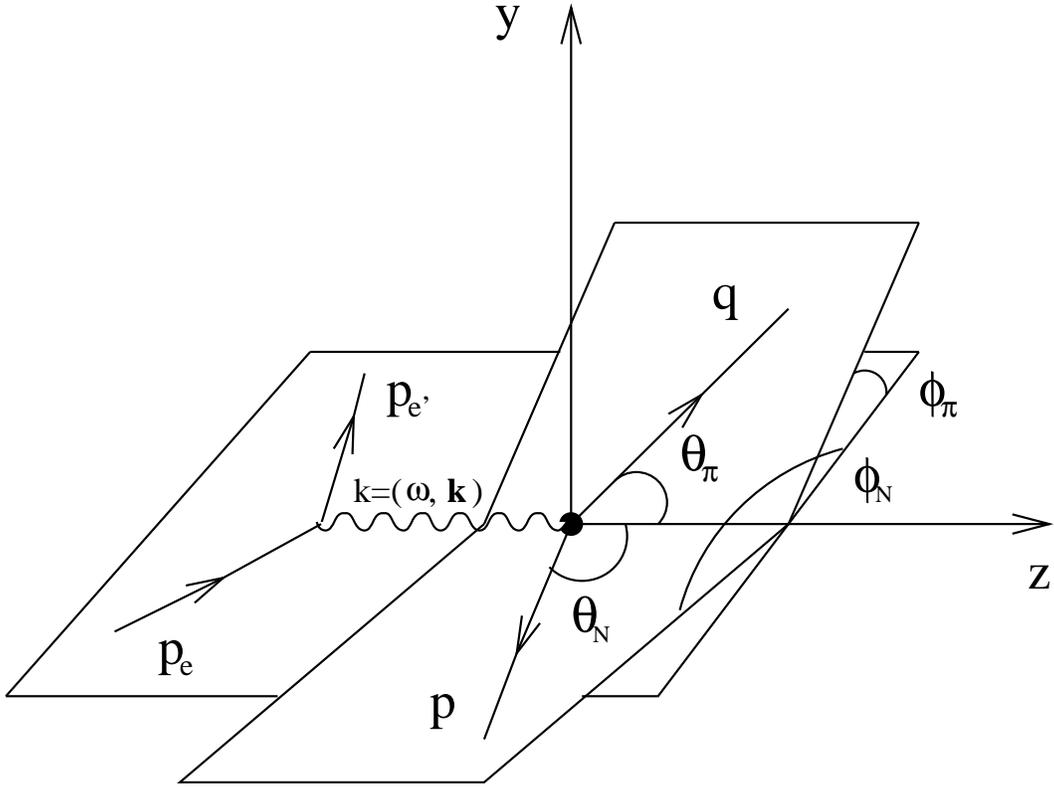,width=14cm}}
\vspace{1cm}
\caption{The coordinate system of the reaction
$A(e,e^\prime \pi N)B$ in the laboratory frame.}
\label{eexyz1}
\end{figure}

In this section, we define our coordinate system and discuss some
kinematic aspects of the exclusive reaction $(e,e^\prime\pi N)$
on nuclei.  In the laboratory frame, the four-momentum of
the incoming electron is $p_e =(E_e , {\bf p}_e)$,
of the outgoing electron $p_{e^{\prime}}=(E_{e^{\prime}},{\bf
p}_{e^{\prime}})$,
thus the virtual photon carries the four-momentum
$k=p_e -p_{e^{\prime}}=(\omega,{\bf k})$.
The outgoing pion and
nucleon have four-momenta of $q^{\mu}= (E_\pi,{\bf q})$ and
$p^{\mu}= (E_N, {\bf p})$, respectively.
The target nucleus is at rest with mass $M_i$ and the
recoiling residual nucleus of mass $M_f$ has the momentum
\begin{equation}
{\bf Q}={\bf k}-{\bf q}-{\bf p}
\label{mcon}
\end{equation}
 and kinetic energy $T_Q= \frac{Q^2}{2M_f}$.
The momentum transfer ${\bf Q}$ is also referred to as the missing momentum.
 Overall energy conservation requires
\begin{equation}
\omega+M_i = E_\eta +E_N +M_f + T_Q.
\label{econ}
\end{equation}
As shown in Fig.~\ref{eexyz1}, the z-axis is defined by the
virtual photon direction ${\bf k}$,
and we choose the azimuthal angle of the pion, $\phi_\pi = 0$,
by defining ${\bf y} = {\bf k} \times {\bf q}$
and ${\bf x} = {\bf y} \times {\bf z}$.
This puts the x-axis in the electron scattering plane while
the y-axis is normal to the electron scattering plane.
In general, the pion and the nucleon can both go out of the plane.

We assume that the reaction takes place on a single bound nucleon
with four-momentum $p{^\mu}{_i} = (E_i, {\bf p}_i)$ and that
energy and momentum are conserved at this vertex
({\em i.e.}, the impulse approximation).
Thus ${\bf p}_i =-{\bf Q}$ and $E_i=E_\pi +E_N -\omega$.
If ${\bf Q}$ does not vanish, the struck nucleon is off its mass shell.
This is the only reasonable off-shell choice
since the photon, the pion, the outgoing
nucleon, and the recoiling nucleus are all external lines and must be
on their respective mass shells.

The magnitude of the momentum transfer to the recoiling nucleus
has a wide range, including zero, depending on the directions
of the outgoing pion and nucleon with respect to the virtual
photon.  However, since the reaction amplitude is proportional to the
Fourier transform of the bound state single particle wave function,
it becomes quite small for
momentum transfers greater than about 300 MeV/c.
Thus for all but the lightest nuclei we can safely neglect the nuclear
recoil velocity (and recoil energy $T_Q$)
and generate optical potentials for the outgoing
particles in the laboratory frame.  As noted in \cite{li},
since the residual nucleus can take up varying amount of momentum
but little recoil energy, the reaction offers
great kinematic flexibility and by appropriate choices one can
investigate the production operator, the bound state wave-function,
or the final state interaction of the outgoing meson and nucleon.

Another useful kinematic quantity is the invariant mass of
the outgoing $\pi N$ pair $W=\sqrt{(p +q )^2}$
which indicates whether the pion production process takes place
in the $\Delta(1232)$ resonance region.
Note that there is a difference in kinematics between the real and
virtual photon cases. In our case,
the magnitude $|{\bf k}|$ is always larger than $\omega$,
while in the real case they are equal. This results in
different off-shell behaviors of the production amplitudes.

\subsection{Nuclear Structure Inputs}

As discussed in the previous section,
we are primarily interested in cases of low momentum transfer
to the recoiling nucleus.  Thus, our choice
of single particle wave-functions for the bound state is not critical
as long as the basic size of the orbital is described correctly.  For
convenience we use harmonic oscillator wave-functions which have the
advantage that their Fourier transforms are simply obtained.  For each
nucleus under consideration, we adjust the harmonic oscillator
range parameter $b$ until the RMS radius of the ground state charge
distribution agrees with the experimentally determined values.

For the continuum nucleon wave-functions we solve the
Schr\"{o}dinger equation with an optical potential present whenever the
particle is to be detected.  Further, we use the experimentally
determined separation energies for a given orbital in order to fix the
value of the mass of the recoiling nucleus $M_f$.
Many optical models for the outgoing nucleon are available.  We use a
non-relativistic reduction of the global optical model of Clark and
collaborators~\cite{osu}.  This model has the advantage that it fits
nucleon scattering over a wide range of energy and A values,
and hence is very useful for making surveys of a wide range of possible
experimental situations. Other models have been tried~\cite{schw} and
were found to make little difference.
Once experimental data is available for the exclusive
reaction, an optical model specific to the nucleus and energy range of
the outgoing nucleon can be substituted.

For the continuum pion wave-functions, the SMC optical model~\cite{smc}
is used. Alternatively, we have tried the pion global optical
potential in~\cite{kamalov} which is based on multiple scattering
and found that the two approaches give similar results.

Another nuclear structure input is the overlap of the residual nucleus
with the A-1 spectator particles which
just provides an overall normalization, or spectroscopic factor, to the
cross section. These have been taken from quasi-elastic electron
scattering studies.

\subsection{Differential Cross Sections}

The differential cross section for $A(e,e^{\prime}\pi N)B$
in the laboratory frame can be written as
\begin{eqnarray}
d\sigma &=& \frac{E_e}{|{\bf p}_e|}\; \frac{m_e}{E_e}\;
\frac{m_e\,d{\bf p}_{e^{\prime}}}{E_{e^{\prime}}\,(2\pi)^3}\;
\frac{d{\bf q}}{2E_{\pi}\,(2\pi)^3}\;
\frac{m_{\scriptscriptstyle N}\,d{\bf p}}{E_N\,(2\pi)^3}\;
\frac{M_f\,d{\bf Q}}{E_f\,(2\pi)^3} \nonumber \\
& & \times \;\overline{\sum} |M_{fi}|^2\;
 (2\pi)^4\,\delta^{(4)}\,(p_e +P_i -p_{e^{\prime}} -P_f -q -p)
\end{eqnarray}
where $\overline{\sum}$ denotes the sum over final spins and the average
over initial spins and the $\delta$-function defines the overall
energy-momentum conservation.
 After integrating
over ${\bf Q}$ and $|{\bf p}|$ and summing
over the nuclear part of the transition matrix element,
the 5-fold coincidence differential cross section can be written
as a sum over single particle matrix elements times some
kinematic phase space factor:
\begin{eqnarray}
\frac{d^5 \sigma}{dE_{e^{\prime}}\,d\Omega_{e^{\prime}}\,
dE_{\pi}\, d\Omega_{\pi}\,d\Omega_N} & = &
\frac{|{\bf p}_{e^{\prime}}|\,m^2_e}{|{\bf p}_e|\,(2\pi)^3}\;
\frac{M_f m_N\; |{\bf q}|\;|{\bf p}|}
{2(2\pi)^5 |E_N +E_f -E_N \,{\bf p}\cdot({\bf k}-{\bf q})/p^2|} \nonumber \\
& &  \times \;
\frac{1}{2} \sum_{\scriptstyle s\,s^{\prime} \atop\scriptstyle
m_s\,\alpha} \frac{S_{\alpha}}{2j+1} |T(s,s^{\prime},m_s, \alpha)|^2.
\label{5fold}
\end{eqnarray}
Here, $\alpha=\{nljm\}$ is
the quantum number of the bound nucleon, $s$ and $s^\prime$ are
electron spins, $m_s$ is the spin projection
of the outgoing nucleon and $S_{\alpha}$ is the spectroscopic factor.
The single particle transition matrix element is given by
\begin{equation}
T(s,s^{\prime},m_s,\alpha ) =\int d^3 r\,
\Psi^{(+)}_{m_s}({\bf r},-{\bf p})\;
\phi^{(+)}_{\pi}({\bf r},-{\bf q})\;
j_{\mu}\frac{-e}{k^2}J^{\mu}\;
e^{i{\bf k}\cdot{\bf r}}\;
\Psi_{\alpha}({\bf r})
\end{equation}
In this equation,
$j_{\mu}=\bar{u}(p_{e^{\prime}}, s^{\prime})\,\gamma_{\mu}\,u(p_e,
s)$ is the electron current, e the electron charge,
$\Psi_{\alpha}$ the bound nucleon wave-function,
$\Psi^{(+)}_{m_s}$, $\phi^{(+)}_{\pi}$ the distorted outgoing
nucleon and pion wave-functions,
and $J^\mu$ the hadronic transition current.
Introducing an auxiliary current
\begin{equation}
a_{\mu}=j_{\mu}-\frac{j_0}{\omega}k_{\mu} \label{transform}
\end{equation}
and using the fact $k_{\mu}J^{\mu}=0$ and $a_0=0$, one can
eliminate the time component of the hadronic current and
work with the space components only, i.e., $j_{\mu}J^{\mu}=-a_i J_i
=-a_i \delta_{ij} J_j=-\sum_{\lambda}a^i\varepsilon_i(\lambda)
\,\varepsilon_j(\lambda)J_j$ where we have inserted a dyadic
to project the expression onto the virtual photon polarization
basis which is chosen to be unit vectors along the three
coordinate axes ($\lambda=x,y,z$).
Under this basis, the matrix element squared
with sums over the electron spins can be written as
\begin{equation}
\sum_{s\,s^{\prime}}|T|^2 =\frac{e^2}{k^4}
\sum_{\lambda\,\lambda^{\prime}}
\rho_{\scriptscriptstyle \lambda\,\lambda^{\prime}} \;
w^{\ast}_{\scriptscriptstyle \lambda}
w_{\scriptscriptstyle\lambda^{\prime}}  \label{t2}
\end{equation}
where we have defined the electron density matrix elements
\begin{equation}
\rho_{\scriptscriptstyle \lambda\,\lambda^{\prime}}=\sum_{s s^{\prime}}
\left[\mbox{\boldmath $\varepsilon$}(\lambda)\cdot{\bf a}\right]^{\ast}
\;\left[\mbox{\boldmath $\varepsilon$}(\lambda^\prime)
\cdot{\bf a}\right]  \label{rho}
\end{equation}
and the hadronic matrix elements
\begin{equation}
w_{\scriptscriptstyle \lambda}(m_s, \alpha) = \int d^3 r\,
\Psi^{(+)}_{m_s}({\bf r},-{\bf p})\;
\phi^{(+)}_{\pi}({\bf r},-{\bf q})\;
\mbox{\boldmath $\varepsilon$}(\lambda)\cdot {\bf J}\;
e^{i{\bf k}\cdot{\bf r}}\;
\Psi_{\alpha}({\bf r}).  \label{w}
\end{equation}
The matrix element
$w_{\scriptscriptstyle \lambda}(m_s,\alpha)$ has
the identical form to that in the real photon case
(see Eq.~(6) of Ref.~\cite{li}) except here one has the
longitudinal polarization in addition to the two transverse polarizations.
Therefore, the method for evaluating Eq.~(\ref{w})
closely follows Ref.~\cite{li} and we will not be repeated here.

After carrying out the electron spin sums, and with the help of
Eq.~(\ref{k2}) and the relations
\begin{eqnarray}
& & p^x_e=p^x_{e^{\prime}}=\frac{|{\bf p}_e|\,|{\bf p}_{e^{\prime}}|}
{|{\bf k}|}\, \sin \theta_e, \;\;\;\;\;
p^y_e=p^y_{e^{\prime}}=0, \nonumber \\
& & p^z_e=\frac{|{\bf p}_e|}{|{\bf k}|}\,(|{\bf p}_e|-
|{\bf p}_{e^{\prime}}|\,\cos\theta_e ), \;\;\;
 p^z_{e^{\prime}}=\frac{|{\bf p}_{e^{\prime}}|}{|{\bf k}|}\,
(-|{\bf p}_{e^{\prime}}|+|{\bf p}_e|\,\cos\theta_e ),
\end{eqnarray}
we find the $3\times 3$ electron density matrix
\begin{equation}
\rho=\frac{-k^2}{m_e^2}\;\frac{1}{1-\epsilon}
\pmatrix{{1\over2}(1+\epsilon)&0&-{1\over2}
\sqrt{\frac{-2k^2}{\omega^2}\epsilon\, (1+\epsilon)}\cr
0&{1\over2}(1-\epsilon)&0\cr
-{1\over2}\sqrt{\frac{-2k^2}{\omega^2}\epsilon\, (1+\epsilon)}
&0& \frac{-k^2}{\omega^2}\epsilon \cr}, \label{matrix}
\end{equation}
where $\epsilon$ has been defined in Eq.~(\ref{eps}).
Using this matrix, the 5-fold differential cross section
 Eq.~(\ref{5fold})
can be cast into the form of an electron flux factor times
the virtual photon cross section, just as in the single nucleon
case:
\begin{equation}
\frac{d^5 \sigma}{dE_{e^{\prime}}\,d\Omega_{e^{\prime}}\,
dE_{\pi}\, d\Omega_{\pi}\,d\Omega_N} =
\Gamma\;\frac{d^3 \sigma_v}{dE_{\pi}\, d\Omega_{\pi}\,d\Omega_N},
\label{5fold1}
\end{equation}
except that the electron flux factor here is defined by
\begin{equation}
\Gamma = \frac{\alpha}{2{\pi}^2}\,\frac{E_{e^{\prime}}}{E_e}\,
\frac{E_{\gamma}}{(-k^2)}\,\frac{1}{1-\epsilon}
\end{equation}
where $E_\gamma =(s^2 -M_i^2)/{2M_i}$, and $s=k+P_i$.
The virtual photon cross section is given by
\begin{equation}
\frac{d^3 \sigma_v}{dE_{\pi}\, d\Omega_{\pi}\,d\Omega_N}
= d^3 \sigma_{\scriptscriptstyle T}
+\epsilon \, d^3 \sigma_{\scriptscriptstyle L}
+\epsilon\,d^3 \sigma_{\scriptscriptstyle TT}
+\sqrt{\epsilon (1+\epsilon)}\,d^3 \sigma_{\scriptscriptstyle TL},
\label{virtual}
\end{equation}
where we have introduced
the short-hand notations for the transverse cross section:
\begin{equation}
d^3 \sigma_{\scriptscriptstyle T} \equiv \frac{d^3 \sigma_{\scriptscriptstyle
T}}{dE_{\pi}\, d\Omega_{\pi}\,d\Omega_N}
=C\,\sum_{\alpha\,m_s}\;\frac{S_{\alpha}}{2j+1}\;
\frac{1}{2} \left( |w_x|^2+|w_y|^2 \right),
\end{equation}
the longitudinal cross section:
\begin{equation}
d^3 \sigma_{\scriptscriptstyle L} \equiv \frac{d^3 \sigma_{\scriptscriptstyle
L}}{dE_{\pi}\, d\Omega_{\pi}\,d\Omega_N}
=C\,\sum_{\alpha,m_s}\;\frac{S_{\alpha}}{2j+1}\;
\frac{-k^2}{\omega^2}\,|w_z|^2
\end{equation}
the polarization cross section:
\begin{equation}
d^3 \sigma_{\scriptscriptstyle TT} \equiv \frac{d^3 \sigma_{\scriptscriptstyle
TT}}{dE_{\pi}\,
 d\Omega_{\pi}\,d\Omega_N}
=C\,\sum_{\alpha\,m_s}\;\frac{S_{\alpha}}{2j+1}\;
\frac{1}{2} \left( |w_x|^2-|w_y|^2 \right),
\end{equation}
and the interference cross section:
\begin{equation}
d^3 \sigma_{\scriptscriptstyle TL} \equiv \frac{d^3 \sigma_{\scriptscriptstyle
TL}}{dE_{\pi}\,
 d\Omega_{\pi}\,d\Omega_N}
=C\,\sum_{\alpha\,m_s}\;\frac{S_{\alpha}}{2j+1}\;
\frac{-1}{2}\,\sqrt{\frac{-2k^2}{\omega^2}}\,
\left( w_x^{\ast} w_z + w_z^{\ast} w_x \right).
\end{equation}
The kinematic factor C is given by
\begin{equation}
C=\frac{M_f m_N\; |{\bf q}|\;|{\bf p}|}
{4(2\pi)^5 E_{\gamma}|E_N +E_f -E_N \,
{\bf p}\cdot({\bf k}-{\bf q})/p^2|}.
\end{equation}
The above definitions of the factors $\Gamma$ and C ensure that
in the real photon limit $-k^2\rightarrow 0$,
the transverse cross section $d^3\sigma_{\scriptscriptstyle T}$
reduces to the real photon cross section (see Eq.~(3) of Ref.~\cite{li}).

\section{Results}
\label{result}

In this section we present our results for exclusive quasifree
pion electroproduction on nuclei.
There exist almost no data for the reaction, except for one
experiment recently carried out at SLAC~\cite{slac} which
will be discussed below. However, the experiment
$^{16}O(e,e^\prime \pi^- p)^{15}O$ is being planned at the
new high duty cycle electron accelerator at Mainz~\cite{ros}.
Such experiments are difficult to perform since
they require large solid angle acceptance, good energy resolution,
and long beam time.
We expect our results to be helpful in planning such experiments.
We use $^{12}$C as the generic target and consider processes
leaving the residual nucleus in its ground state. This corresponds to
knocking out a p-state nucleon from $^{12}$C.
Our formalism is equipped to study other nuclei as well.
Our goal is to get an overview of the magnitude and shape of the cross sections
and to seek out suitable kinematics for future experiments.
For this particular purpose, it is sufficient to use the plane wave impulse
approximation (PWIA) which costs much less computer time.
The role of distortions will be discussed separately.

We present our results using three kinematic arrangements:\\
I)  the magnitude of the missing momentum is kept fixed (quasifree
kinematics), \\
II) the magnitudes of all momenta are fixed (fixed kinematics), \\
III) the missing momentum ${\bf p}_m$ is allowed to vary freely (open
kinematics). \\
We will consider coplanar setups first, i.e., the pion
plane and the nucleon plane coincide with the electron scattering
plane with $\phi_\pi=0^o$ and $\phi_N=180^0$ (see Fig.~\ref{eexyz1}).
Out of plane arrangements will be presented when we discuss the separation
of cross sections.

\subsection{Results for Kinematics I}

In this kinematic arrangement, we keep  $\omega$, ${\bf k}$,
and the magnitude of the missing momentum $Q$ fixed,
and study the pion angular distribution.
We refer to it as `quasifree kinematics' since it most resembles the free
two-body kinematics. It is achieved in two steps.
For a given pion angle, we set $Q=0$ and calculate the corresponding
`two-body' kinematics inside the nucleus.
Note that this is not exactly the same as the free two-body case
since nuclear binding has to be taken into account.
Next, we select a finite value for $Q$ (or $p_m$), thus violating
momentum conservation (Eq.~(\ref{mcon})). However,
 energy conservation (Eq.~(\ref{econ})) is almost maintained
because the nuclear recoil energy $T_Q$ is small.
As a result, the pion and proton energies remain
at their two-body values for a fixed pion angle,
 but the proton angle changes to a new value.
This setup is suitable for studying effects of final state interactions
since the pion and proton energies both change as the pion angle is
varied over the whole angular range.

\begin{figure}
\centerline{\psfig{file=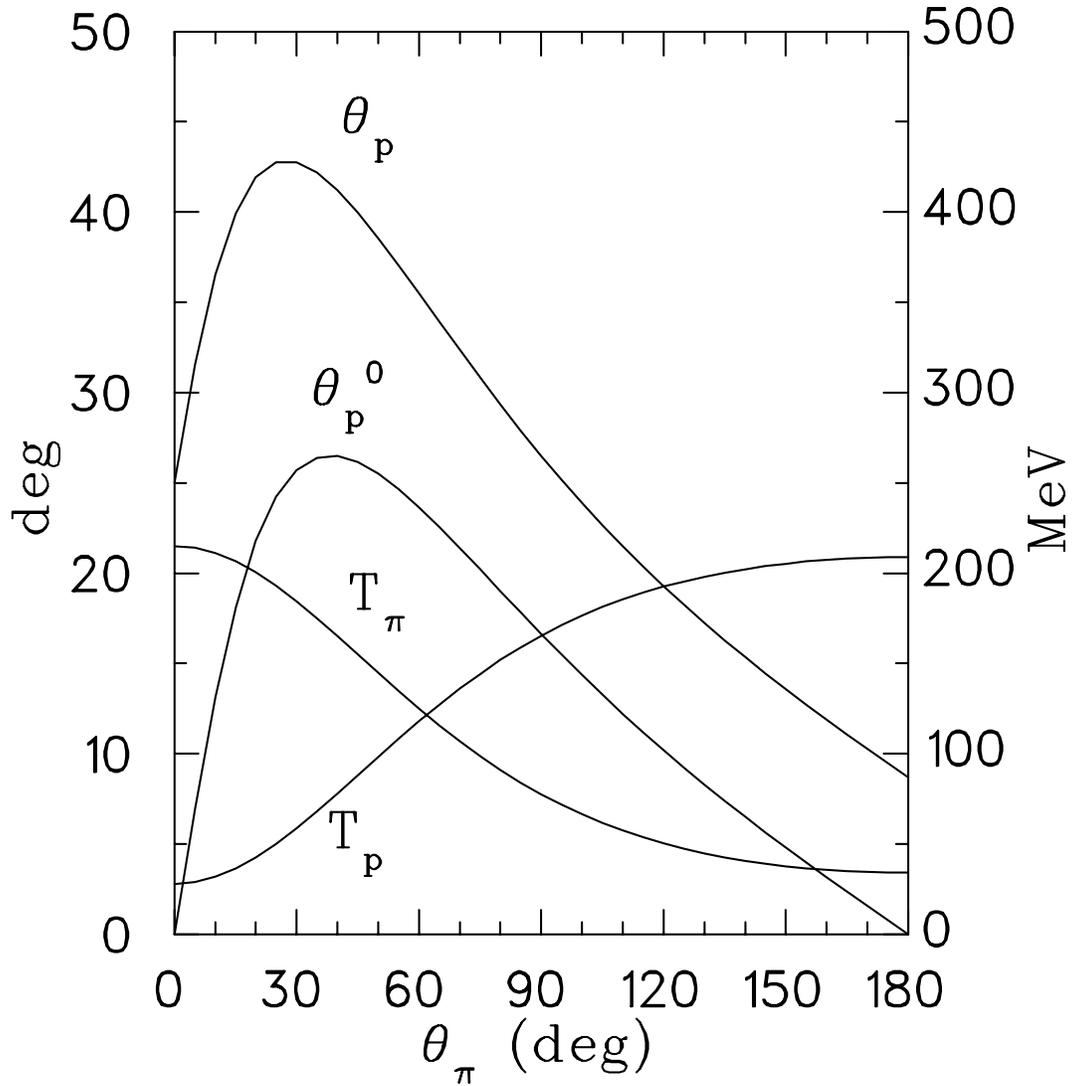,width=14cm}}
\vspace{1cm}
\caption{Kinematic variables as a function of the pion angle
for one case of Kinematics I.
The scale for angles is on the left side
and the scale for energies is on the right side.
The specified variables are: $\omega$=400 MeV, $\epsilon$=0.95,
$k^2=-0.15 (GeV/c)^2$, $\phi_\pi=0$, $\phi_p=180^0$, $Q$=100 MeV/c
and $\theta_\pi$.}
\label{varykin}
\end{figure}

To be more specific, we show in Fig.~\ref{varykin} an example for
the reaction $^{12}C(e,e^\prime\pi^- p)^{11}C_{g.s.}$ which
corresponds to knocking out a neutron from the $p_{3/2}$ orbital.
The solved kinematic variables (the proton angle $\theta^0_p$ at
Q=0, and $\theta_p$ at finite Q, the proton kinetic energy $T_p$, the
pion kinetic energy $T_\pi$) are plotted as a function of the pion
angle. The invariant mass W of the $\pi N$ pair remains roughly
constant around 1210 MeV over the range.
Note that different Q values are necessary to generate relatively
large cross sections for orbitals with different angular momentum
(around zero for s-waves, about 100 MeV/c for p-waves, and
about 150 MeV/c for d-waves, etc).

\begin{figure}
\centerline{\psfig{file=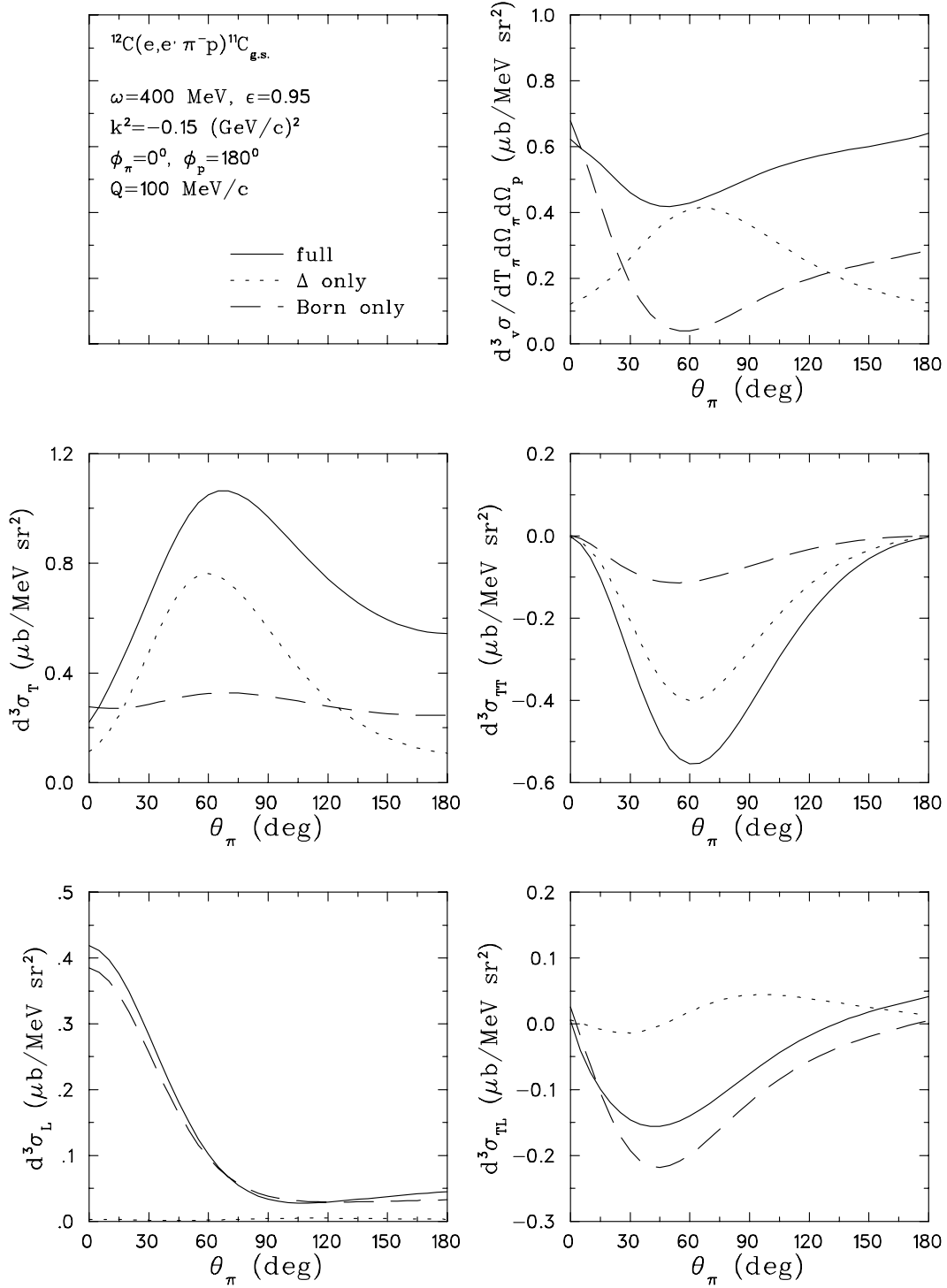,width=14cm}}
\vspace{1cm}
\caption{Pion angular distributions of the
virtual differential cross sections
for $\pi^-$ electroproduction from $^{12}C$ are plotted
under kinematics I,
along with the contributions from the background and the
resonant terms in the production operator. The kinematics and the
legends to the curves are given in the figure.}
\label{vary}
\end{figure}

Fig.~\ref{vary} shows the charged pion virtual cross section along
with the individual structure functions
for such a kinematic setup.
The energy of the virtual photon is chosen to be $\omega=400 MeV$,
which lies in the $\Delta$ region.
The values for $\omega$ and $-k^2$ can be varied separately
by tuning the initial and final electron scattering energies and angles.
One can obtain the full 5-fold differential cross section,
Eq.~(\ref{5fold1}),
by multiplying the virtual cross section with the electron flux factor
which depends on the values of electron scattering energies and angle
actually used in the experiment.
In forward pion direction the cross section is dominated by the
longitudinal structure function. This is similar to the free pion
electroproduction process on the nucleon where the dominance of the
longitudinal term at low $t$ (corresponding to small $\theta_{\pi}$) has
been used to extract the pion electromagnetic form factor in a Chew-Low
extrapolation. Our findings confirm that this situation also occurs in
the nuclear case, possibly permitting the extraction of the pion form
factor in the nuclear medium. At larger pion angles,
$d^3 \sigma_{\scriptscriptstyle L}$
rapidly decreases and
$d^3 \sigma_{\scriptscriptstyle T}$
becomes dominant. The interference and polarization cross sections
contribute moderately for intermediate pion angles with negative sign.
Shown separately are the contributions from the
background Born terms and the resonant $\Delta$ term.
As they add coherently there are large interference effects
between the two contributions. Note that the longitudinal structure
function is completely dominated by the Born terms, more specifically,
the main contribution comes from the pion pole term. The $\Delta$ term
is small for
$d^3 \sigma_{\scriptscriptstyle TL}$ whereas
in $d^3 \sigma_{\scriptscriptstyle TT}$ the Born terms are small. Thus
each structure function can provide complementary information on the
production process in the nuclear medium.

\begin{figure}
\centerline{\psfig{file=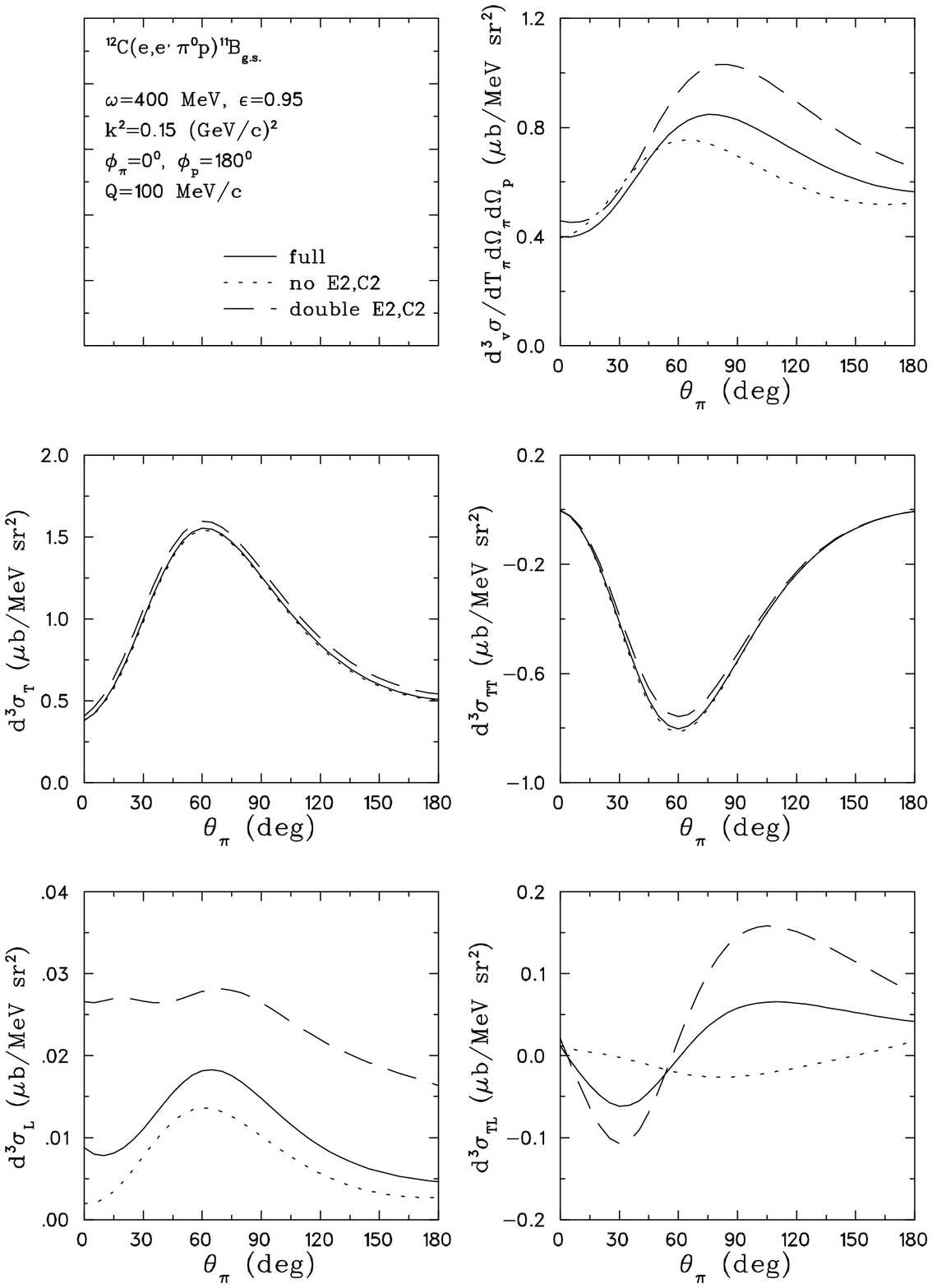,width=14cm}}
\vspace{1cm}
\caption{Sensitivities to the quadrupole component of the $\Delta$ excitation
for $\pi^0$ electroproduction under kinematics I.}
\label{varye2}
\end{figure}

The quadrupole component of the $\Delta$ has been an
important issue in electroproduction processes because it relates
to the internal structure of the $\Delta$ resonance.
Unlike the real photon which can only excite the transverse quadrupole
component E2,
the virtual photon allows for the additional
longitudinal component L2 (or C2).
The quadrupole component is small (only a few percent)
compared to the dominant magnetic dipole transition (M1)
and its value is not well determined.
Fig.~\ref{varye2} shows the
sensitivities of the response functions to E2 and L2 in neutral pion
electroproduction from nuclei by varying the
E2/M1 ratio (see Appendix~\ref{blvirtual}).
We find large sensitivities in the longitudinal and
interference terms while the transverse cross sections are completely
insensitive. This is again similar to the situation on the free nucleon.
Thus, separating these cross sections would allow to verify if the E2/M1
ratio is modified in the nuclear medium.
These sensitivities appear mostly in the $\pi^0$ channel, due to the
small size of the Born terms.

\begin{figure}
\centerline{\psfig{file=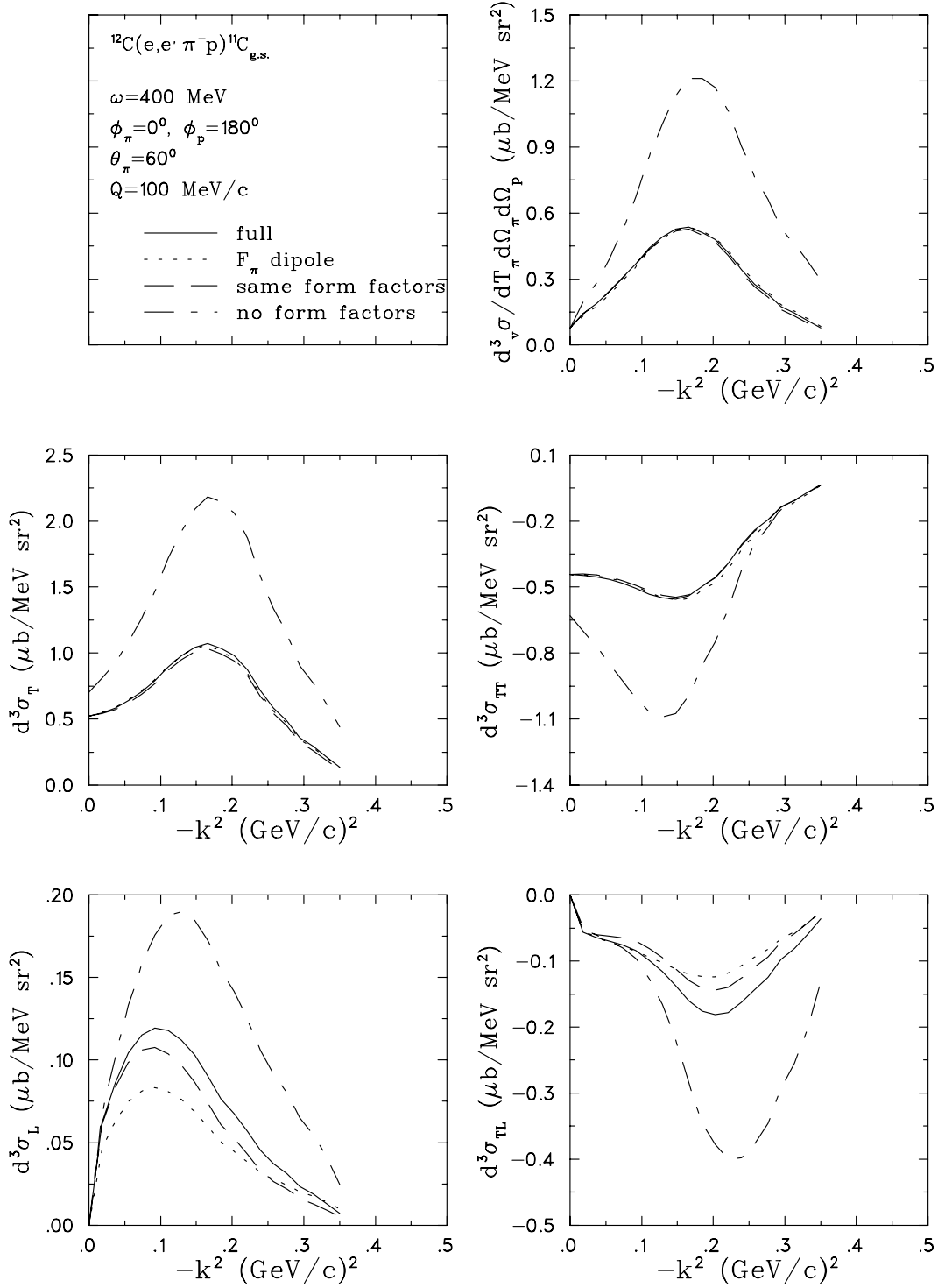,width=14cm}}
\vspace{1cm}
\caption{$k^2$ dependence of the cross sections under kinematics I.
Also shown are the sensitivities to different choice of form factors.}
\label{varyk2}
\end{figure}

Fig.~\ref{varyk2} shows the dependence on the four-momentum
squared, $k^2$, and demonstrates the
sensitivity to form factors. The solid curve displays
the full operator with realistic form factors and gauge fixed, the
short-dashed curve contains a dipole pion form factor instead of a
monopole type,
the long-dashed curve is the result with all form factors equal.
Most sensitivities appear in the longitudinal and the interference
terms. Note, that under the given kinematic conditions
$k^2$ is constrained to below -0.35 $(GeV/c)^2$.

\begin{figure}
\centerline{\psfig{file=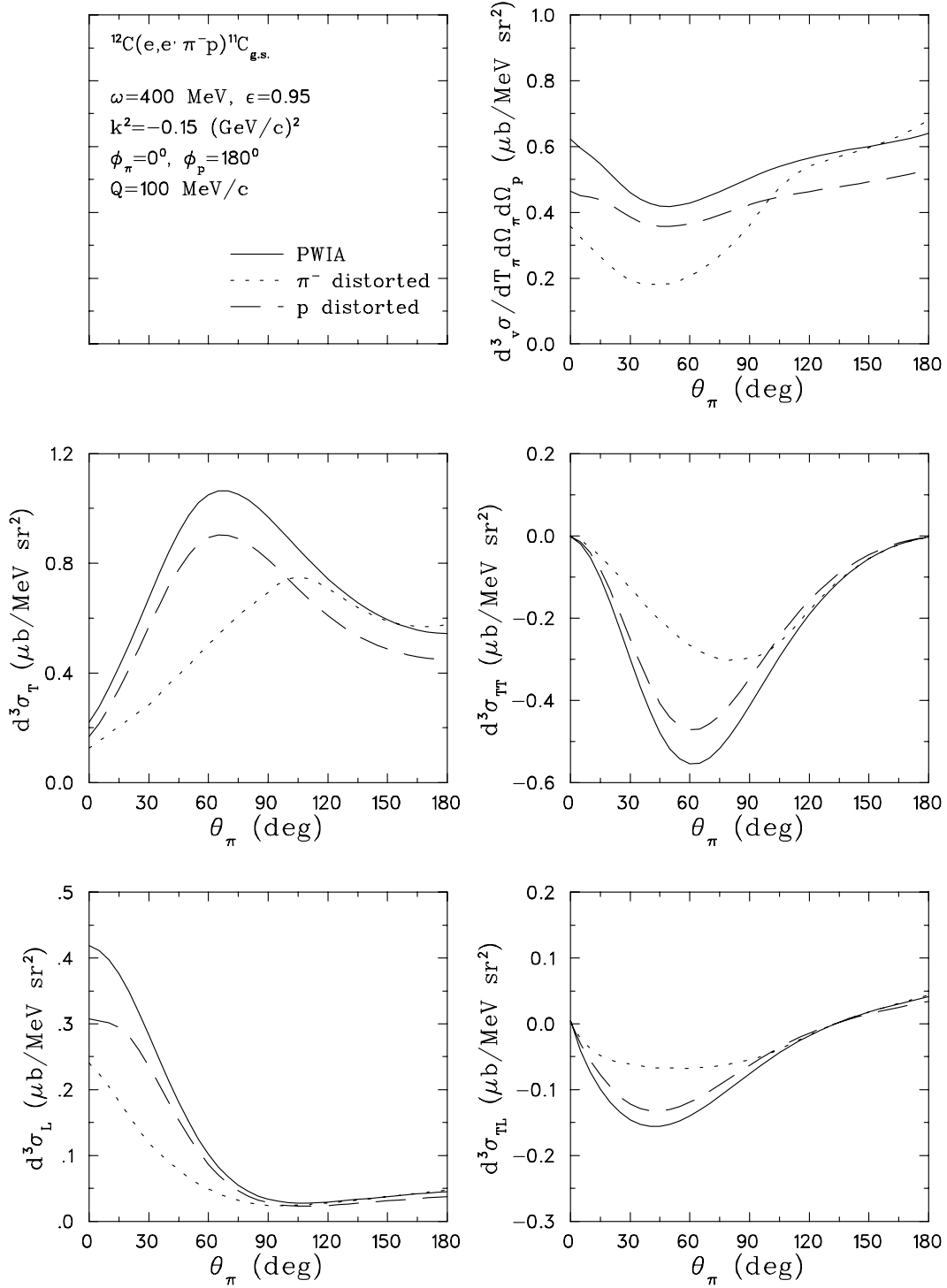,width=14cm}}
\vspace{1cm}
\caption{Distortion effects due to final state interactions
of the pion and the nucleon with the residual nucleus
under kinematics I.}
\label{varydw}
\end{figure}

Fig.~\ref{varydw} presents the
distortion effects due to the final state interactions
of the pion and proton with the residual nucleus.
For details on the distorted waves employed,
see Ref.~\cite{li} and references therein.
One notices the strong energy dependence of the pion distortion since
the pion energies large for small pion angles and vice versa.
The nucleon distortion slightly reduces the transverse cross section as
well as the longitudinal one at forward angles.
The DWIA results given in this figure can serve as a guide on how
distortion affects the PWIA results for kinematics II and III
below since they are given at fixed pion and proton energies.

\subsection{Results for Kinematics II}

In this kinematic arrangement, we keep  $\omega$, ${\bf k}$, the
pion energy $T_\pi$, the nucleon energy $T_N$ and the magnitude $Q$ fixed,
and again vary the pion angle.
We refer to it as the `fixed kinematics'. Only the angles of the
momentum vectors vary.
This has the advantage that the uncertainties from nuclear
structure effects and final-state interactions are minimized,
thus the production process in the medium can be exposed.
Since the lengths of the momentum vectors involved are all fixed,
the angular range accessible is limited.

\begin{figure}
\centerline{\psfig{file=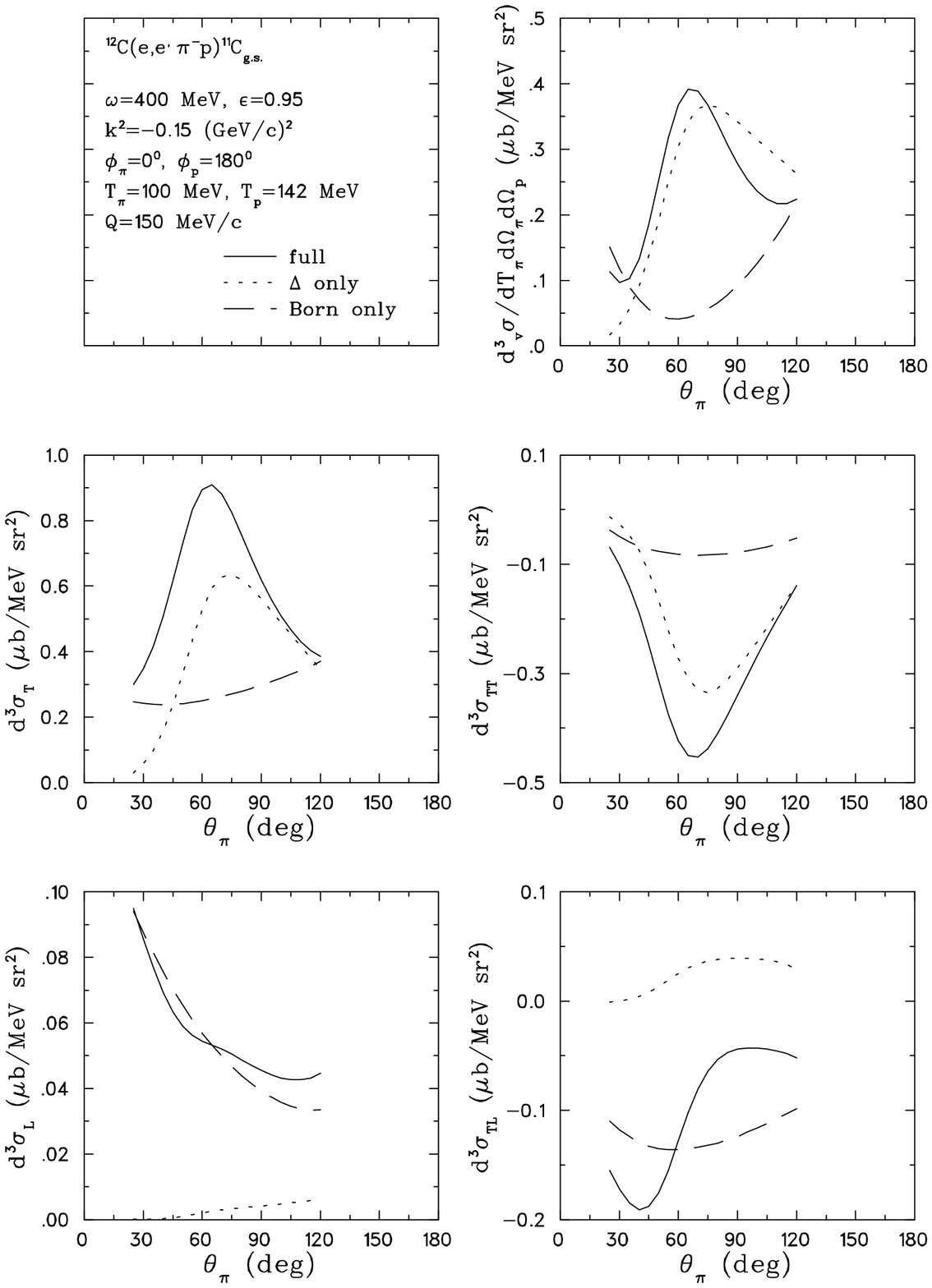,width=14cm}}
\vspace{1cm}
\caption{Pion angular distribution of the
differential cross sections under kinematics II
are plotted along with the contributions from the background and the
resonant terms in the production process.}
\label{fix}
\end{figure}

\begin{figure}
\centerline{\psfig{file=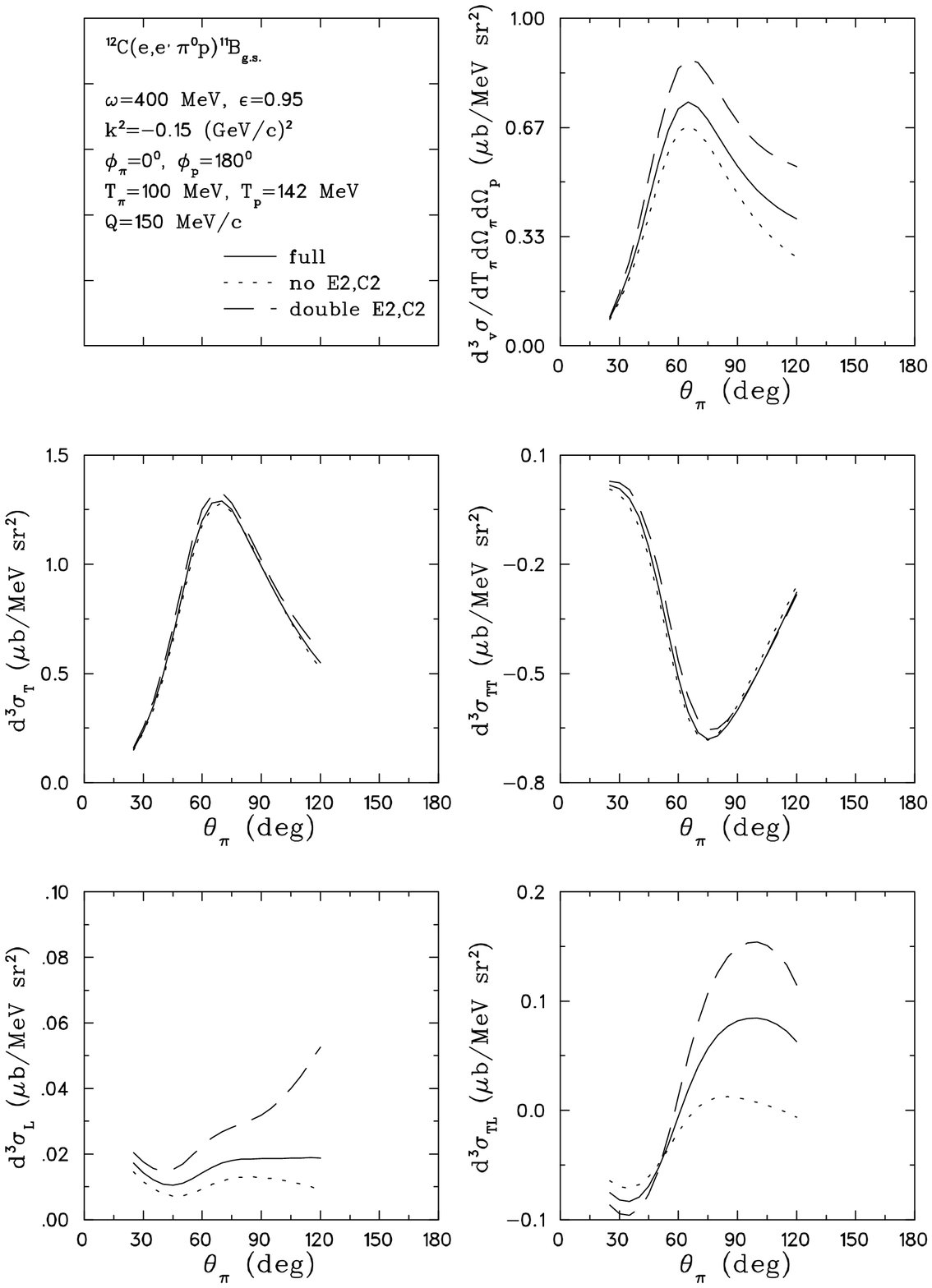,width=14cm}}
\vspace{1cm}
\caption{Sensitivities to the quadrupole component of the
$\Delta$ excitation for $\pi^0$ electroproduction under kinematics II.}
\label{fixe2}
\end{figure}

Fig.~\ref{fix} shows the contributions
from Born and $\Delta$ terms to the charged pion virtual cross
section and the individual response functions as a function of pion angle.
Under the given kinematics, the pion angle covers the range from
$25^0$ to $120^0$.
Fig.~\ref{fixe2} shows the sensitivities to the quadruple moment E2 and L2.

\subsection{Results for Kinematics III}

For this kinematic situation, we keep  $\omega$, ${\bf k}$,
the pion energy $T_\pi$, and nucleon angle $\theta_p$ fixed.
It is equivalent to mapping out the missing momentum ${\bf p}_m$
distribution, we refer to it as the `open kinematics'.
Since ${\bf p}_m$ is allowed to vary, the whole kinematic region is
opened up, although for large $Q$
the cross sections fall off dramatically due to the bound nucleon
wave-functions.

\begin{figure}
\centerline{\psfig{file=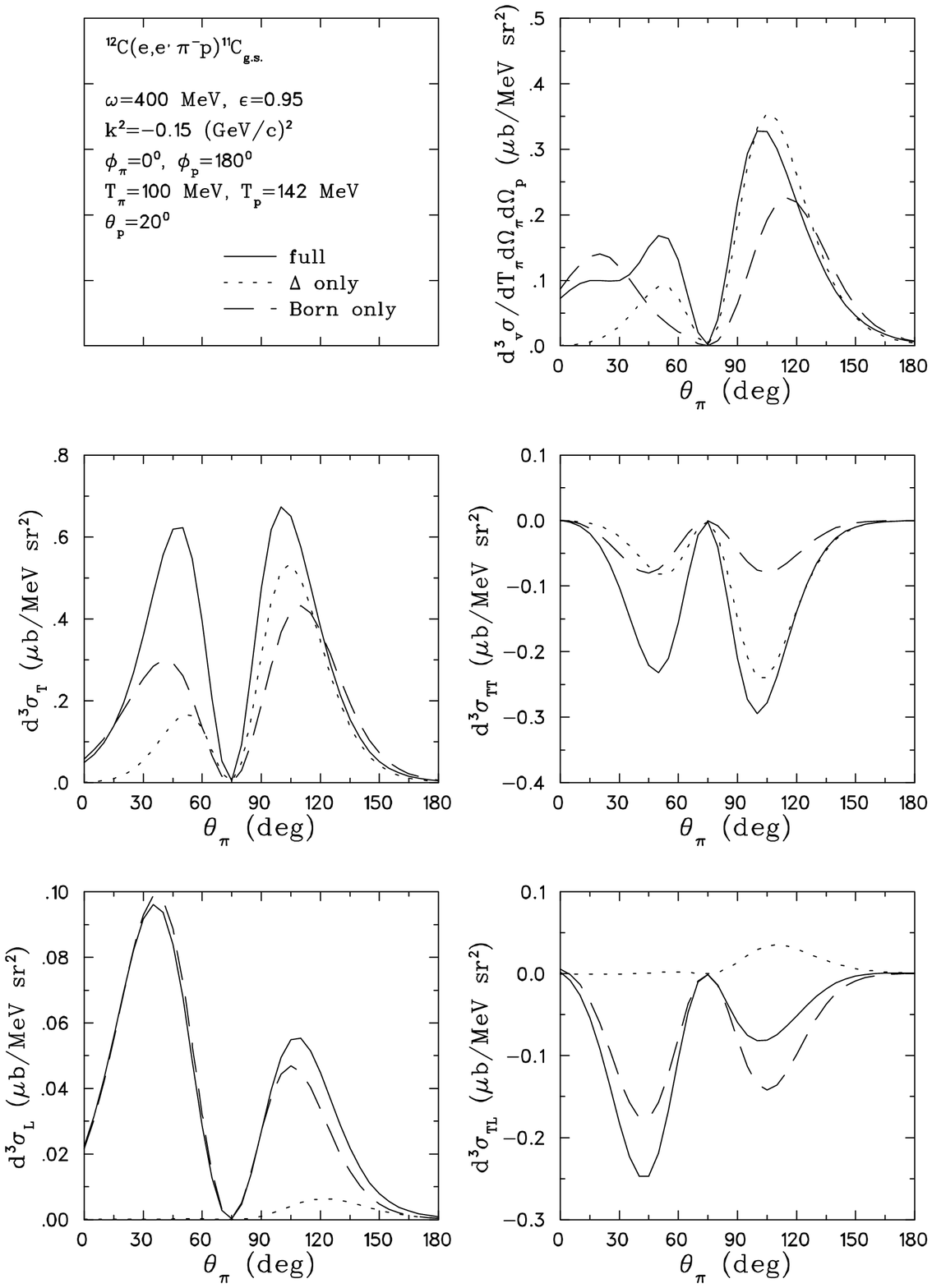,width=14cm}}
\vspace{1cm}
\caption{Pion angular distribution of the
differential cross sections under kinematics III
are plotted along with the contributions from the background and the
resonant terms in the production process.}
\label{pip}
\end{figure}

\begin{figure}
\centerline{\psfig{file=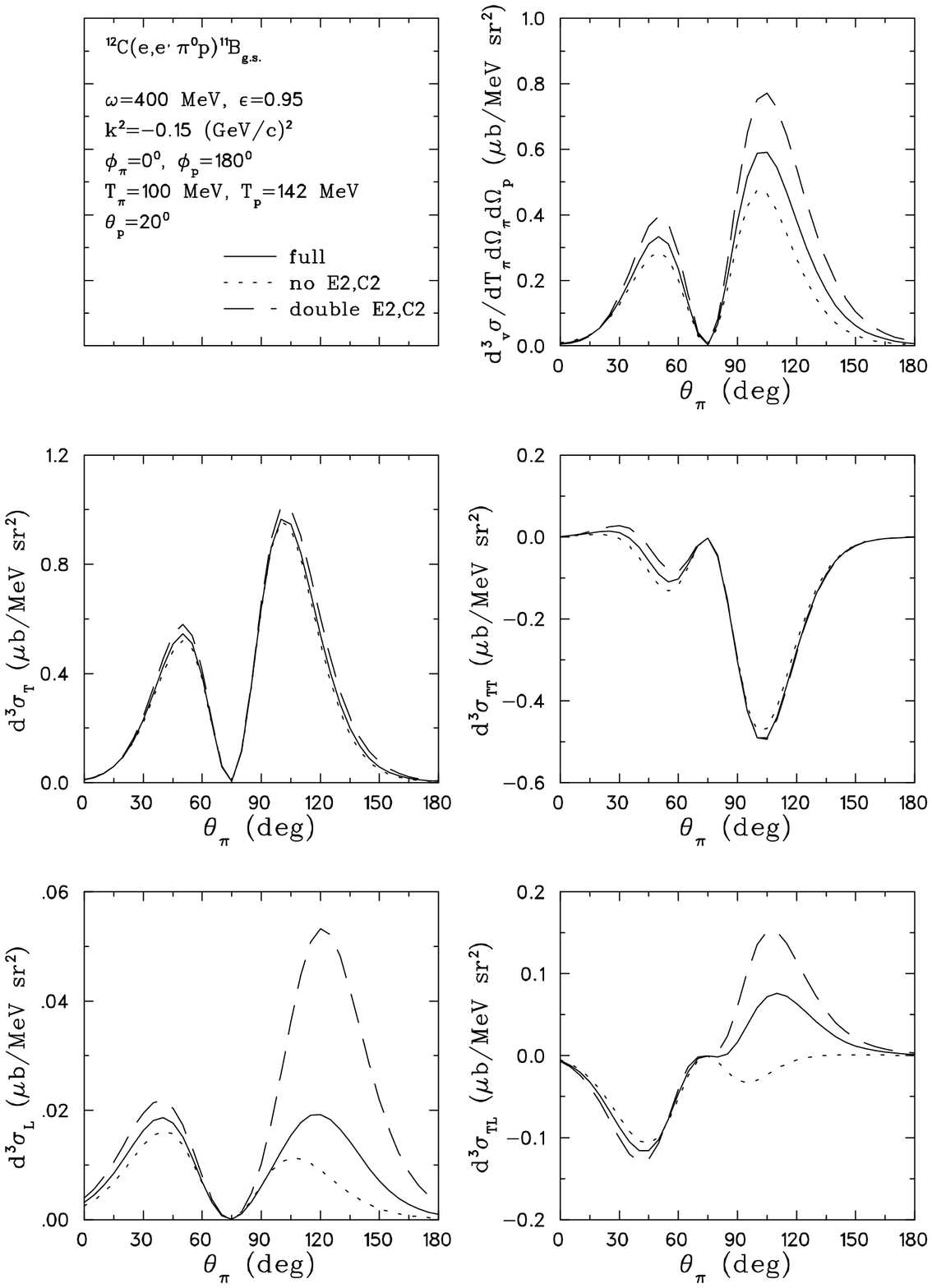,width=14cm}}
\vspace{1cm}
\caption{Sensitivities to the quadrupole component of the
$\Delta$ excitation for $\pi^0$ electroproduction under kinematics III.}
\label{pipe2}
\end{figure}

Fig.~\ref{pip} shows the charged pion angular distributions for one
set of such kinematics.
The twin peaks in the distributions reflect the missing momentum
distribution of the struck nucleon. In this case it is the $p_{3/2}$
state in $^{12}C$.
There is strong interference between the
background and the resonance resulting in a forward-backward
asymmetry of the two contributions in the virtual cross section.
Fig.~\ref{pipe2} shows the sensitivities to the quadrupole component.
Again, large sensitivities are displayed in the longitudinal and the
interference terms.

\begin{figure}
\centerline{\psfig{file=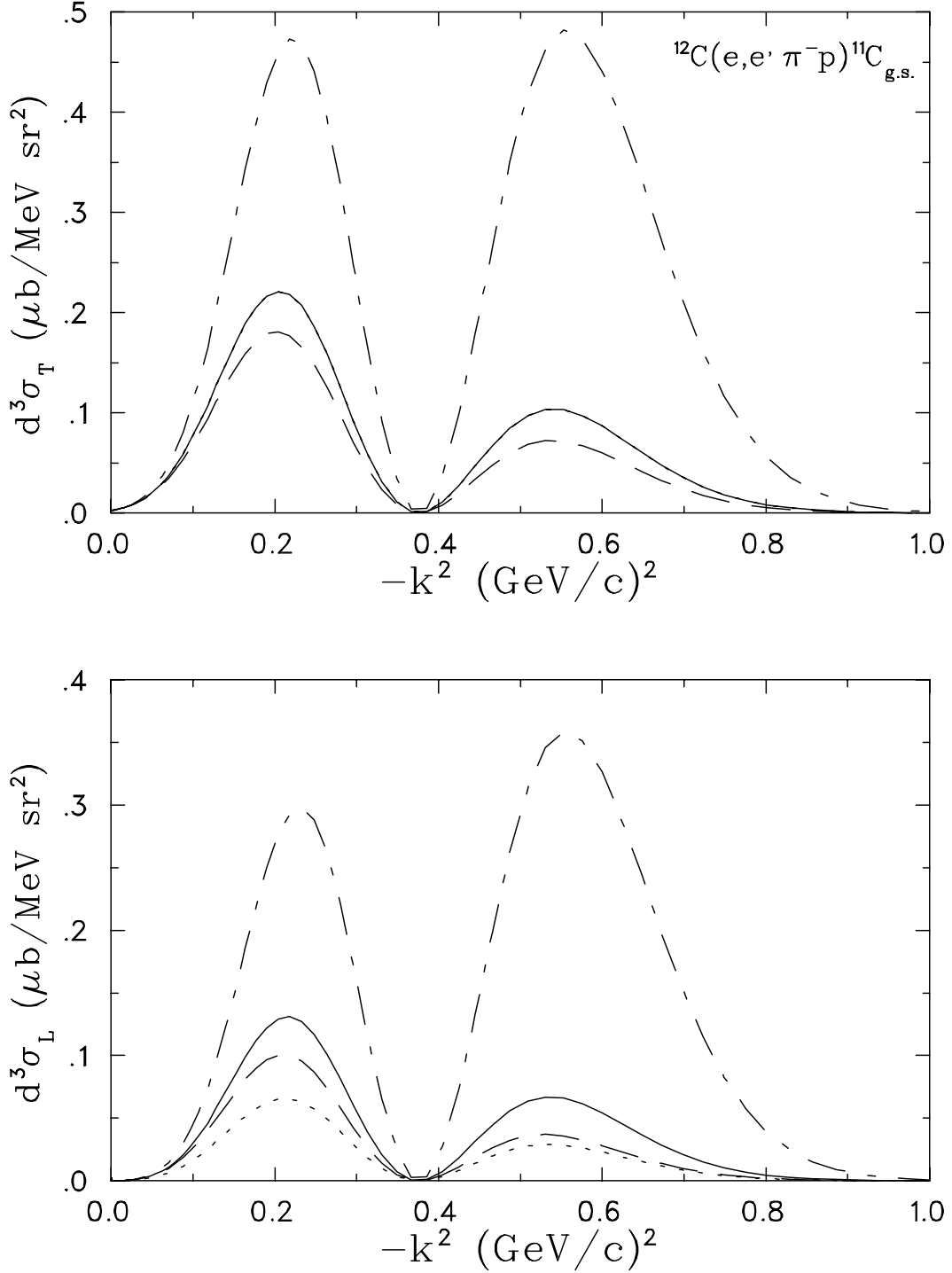,width=14cm}}
\vspace{1cm}
\caption{$k^2$ dependence of the longitudinal and transverse
cross sections in the direction of the virtual photon is shown
under kinematics III along with sensitivities to the form factors.
The full curve is with realistic form factors,
the short-dashed curve is with dipole pion form factor instead of
monopole, the long-dashed curve is with all form factors equal to
$F(k^2)=1/[1+(-k^2)/0.71]^2$, and the dot-dashed curve is with
no form factors.}
\label{piplt}
\end{figure}

Under these kinematic conditions,
when the outgoing pion and proton are both detected along the
direction of the virtual photon, the polarization and the
interference terms vanish.
As a result, the longitudinal and transverse pieces
can be extracted by  a Rosenbluth separation.
Fig.~\ref{piplt} shows the four-momentum $-k^2$
distribution and sensitivities to the form factors.
Large sensitivities to changes in the form factors
 are displayed in the longitudinal term.
Note that in this special case, the longitudinal cross section is
almost as large as the transverse one.
The height of the two peaks are reversed due to the falloff with
the form factors.

\subsection{Separation of differential cross sections}

The virtual cross section contains contributions from the transverse and
longitudinal currents and their interferences, each having different
sensitivities to particular information in the production process.
It is desirable that these individual terms be
disentangled experimentally.
One may recall that in electroproduction
on the nucleon, explicit angular dependences can be factorized
out (see Eq.~(\ref{explicit})), making a separation possible.
It would be desirable to have a similar separation
in pion electroproduction from nuclei.

In a series of studies on quasi-elastic electron scattering off
nuclei~\cite{don}, Donnelly found that angular factors of
the azimuthal angles can indeed be pulled out from the response functions
in the so-called `super Rosenbluth formula'.
Defining the difference
\begin{equation}
\Delta \phi=\phi_\pi-\phi_N
\end{equation}
and the average
\begin{equation}
\phi=\frac{1}{2}\left(\phi_\pi+\phi_N\right),
\end{equation}
It was found that angular factors of the average $\phi$ can be
explicitly separated out, leaving the remainder only dependent
on the difference $\Delta \phi$.
Comparing our formalism to that in Ref.~\cite{don} shows that
the same separation scheme is applicable here.
Translating into our formalism, the angular factors only appear
in the polarization and the interference cross sections:
\begin{equation}
d^3 \sigma_{\scriptscriptstyle TT}=
d^3 \sigma^a_{\scriptscriptstyle TT}\,\cos{2\phi}
+d^3 \sigma^b_{\scriptscriptstyle TT}\,\sin{2\phi},
\end{equation}
\begin{equation}
d^3 \sigma_{\scriptscriptstyle TL}=
d^3 \sigma^a_{\scriptscriptstyle TL}\,\cos{\phi}
+d^3 \sigma^b_{\scriptscriptstyle TL}\,\sin{\phi}.
\end{equation}
where each term is divided into two
terms (with superscripts $a$ and $b$)
that depend only on the difference $\Delta \phi$ multiplied
by the appropriate angular factors that only depend on the
average $\phi$.

Mapping out a $\phi$ distribution
while keeping $\Delta \phi$ fixed allows separating
$(d^3 \sigma_{\scriptscriptstyle T}+\epsilon
d^3 \sigma_{\scriptscriptstyle L})$,
$d^3 \sigma_{\scriptscriptstyle TT}$
and $d^3 \sigma_{\scriptscriptstyle TL}$.
Then, one can use the electron scattering angle $\theta_e$ dependence
in $\epsilon$ to perform a standard Rosenbluth separation of
$d^3 \sigma_{\scriptscriptstyle T}$
and $d^3 \sigma_{\scriptscriptstyle L}$.
Thus, all cross sections in Eq.~(\ref{virtual})
are in principle experimentally accessible.

\begin{figure}
\centerline{\psfig{file=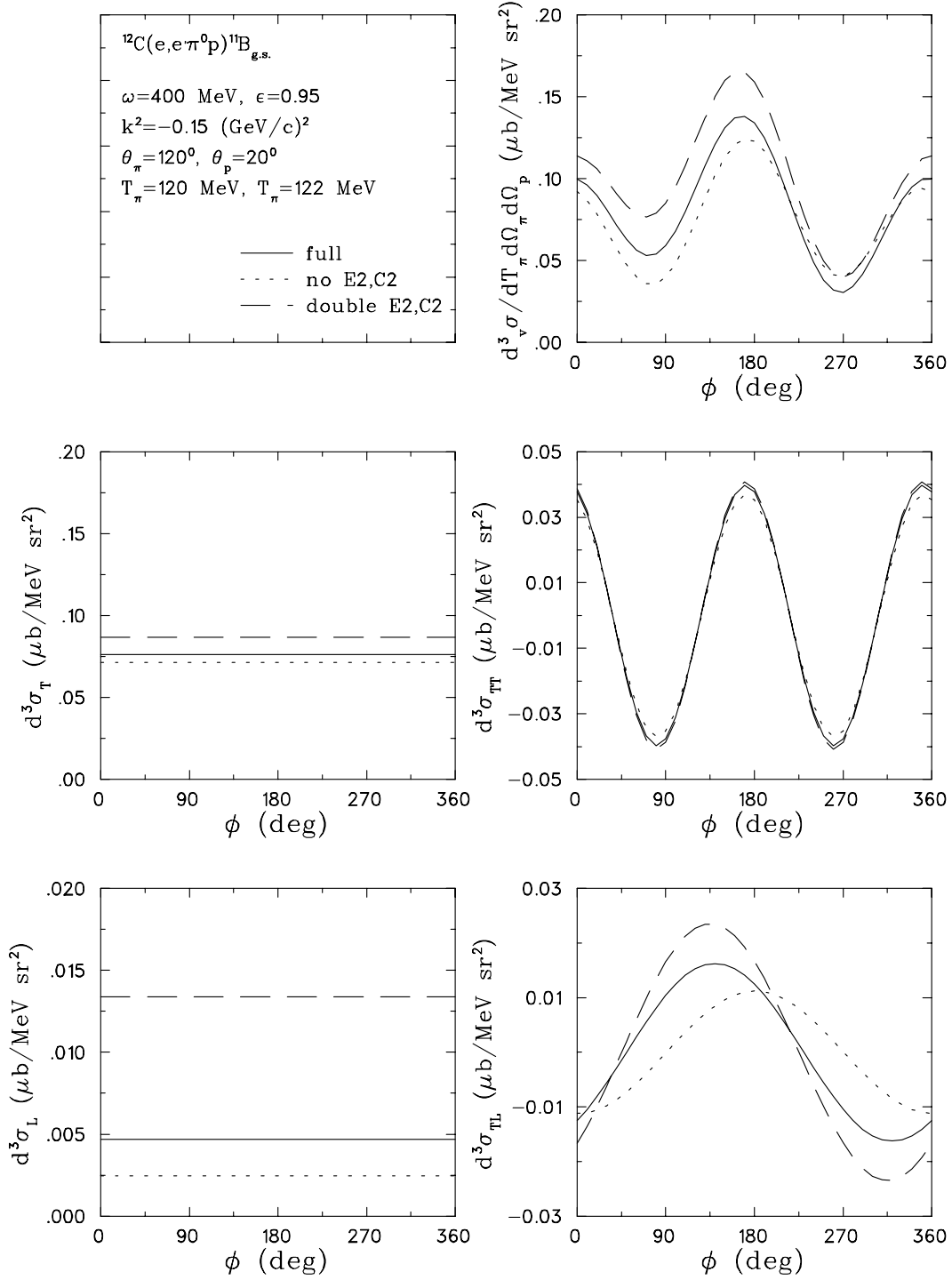,width=14cm}}
\vspace{1cm}
\caption{Out-of-plane distributions of the differential
cross sections are shown as a function of the average azimuthal
angle $\phi=(\phi_\pi+\phi_p)/2$
at fixed $\Delta \phi=\phi_\pi-\phi_p=135^0$ using
kinematics III. Also shown are the sensitivities to the quadruple
moment.}
\label{ph135}
\end{figure}

\begin{figure}
\centerline{\psfig{file=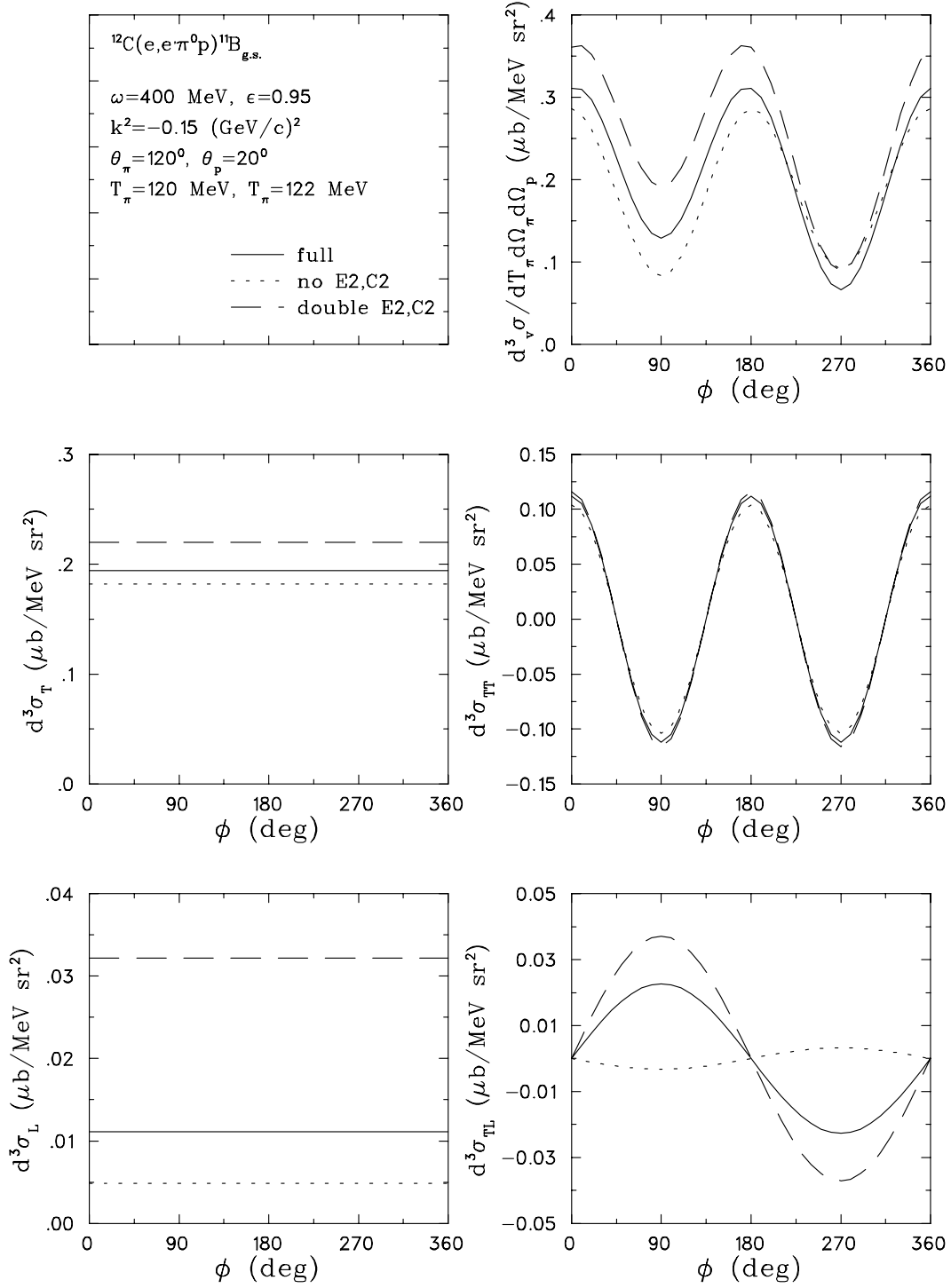,width=14cm}}
\vspace{1cm}
\caption{Same as in Fig.~\protect\ref{ph135},
but at fixed $\Delta \phi=\phi_\pi-\phi_p=180^0$.}
\label{ph180}
\end{figure}

Fig.~\ref{ph135} shows the out-of-plane distribution at
$\Delta \phi=135^0$ using kinematics III.
Also shown are the sensitivities to the quadruple moment.
Note that the polarization cross section
$d^3 \sigma_{\scriptscriptstyle TT}$ is very close to
a pure $\cos{2\phi}$ distribution, indicating that
$d^3 \sigma^b_{\scriptscriptstyle TT}$ is small compared to
$d^3 \sigma^a_{\scriptscriptstyle TT}$. Furthermore, the is no $\phi$
dependence in the longitudinal and transverse structure functions. The
interference term closely resembles a $\sin \phi$ function, thus the
$d^3 \sigma^a_{\scriptscriptstyle TL}$ term is small.
We found that cross sections drop considerably
for configurations of $\Delta \phi< 90^0$ due to increasing
missing momentum. Thus,  $\Delta \phi> 90^0$ configurations are
kinematically favored.
In particular, at $\Delta \phi=180^0$, the polarization cross section
is purely $\cos{2\phi}$ and the interference cross section is purely
$\sin{\phi}$, as shown in Fig~\ref{ph180}.
At $\Delta \phi=0^0$, $d^3 \sigma_{\scriptscriptstyle
TT}$ is purely $\cos{2\phi}$,  and
$d^3 \sigma_{\scriptscriptstyle TL}$ is purely $\cos{\phi}$,
but the cross sections are several orders of
magnitude smaller compared to those at $\Delta \phi=180^0$.

\subsection{Comparison to SLAC data on $A(e,e^\prime p \pi^-)$}

In a recent SLAC experiment~\cite{slac} first measurements were made for
the reactions $(e,e^\prime \pi^- p)$  and $(e,e^\prime p p)$
in the $\Delta$ region on $^2$H, CO, Ar and Xe targets.
The photon-nucleon invariant
mass W ranged from 1.1 to 2 GeV, and $-k^2$ from 0.1 to about 1
$\mbox{GeV/c}^2$,
with an average $-k^2 \approx$ 0.35 $\mbox{GeV/c}^2$.
However, due to low statistics, the exclusive differential cross section
was integrated over the energy and solid angle of the outgoing proton and
over the azimuthal angle of the outgoing pion. In order to compare with
the data, we need to carry out the following integration numerically
\begin{equation}
\frac{d^3\sigma}{dE_{e^\prime}\,d\Omega_{e^\prime}\,d\cos{\theta_\pi}}=
\int \left(
\frac{d^5 \sigma}{dE_{e^{\prime}}\,d\Omega_{e^{\prime}}\,
dE_{\pi}\, d\Omega_{\pi}\,d\Omega_N} \right)\,
dE_\pi\, d\Omega_N\, d\phi_\pi.
\label{pwia.slac}
\end{equation}
This integration can be done in PWIA, but becomes immensely time
consuming in term of computer power for the full DWIA.
So far, we approximate the DWIA result by scaling the PWIA result
with
a distortion factor evaluated at free production kinematics (which
means the missing momentum is zero):
\begin{equation}
\left(\frac{d^3\sigma}{dE_{e^\prime}\,d\Omega_{e^\prime}\,d\cos{\theta_\pi}}
\right)_{DWIA} = \left(\frac{d^5\sigma_{\scriptscriptstyle DWIA}}
{d^5\sigma_{\scriptscriptstyle PWIA}}\right)_{Q=0}
\times \left(
\frac{d^3\sigma}{dE_{e^\prime}\,d\Omega_{e^\prime}\,d\cos{\theta_\pi}}
\right)_{PWIA}
\label{dwia.slac}
\end{equation}

\begin{figure}
\centerline{\psfig{file=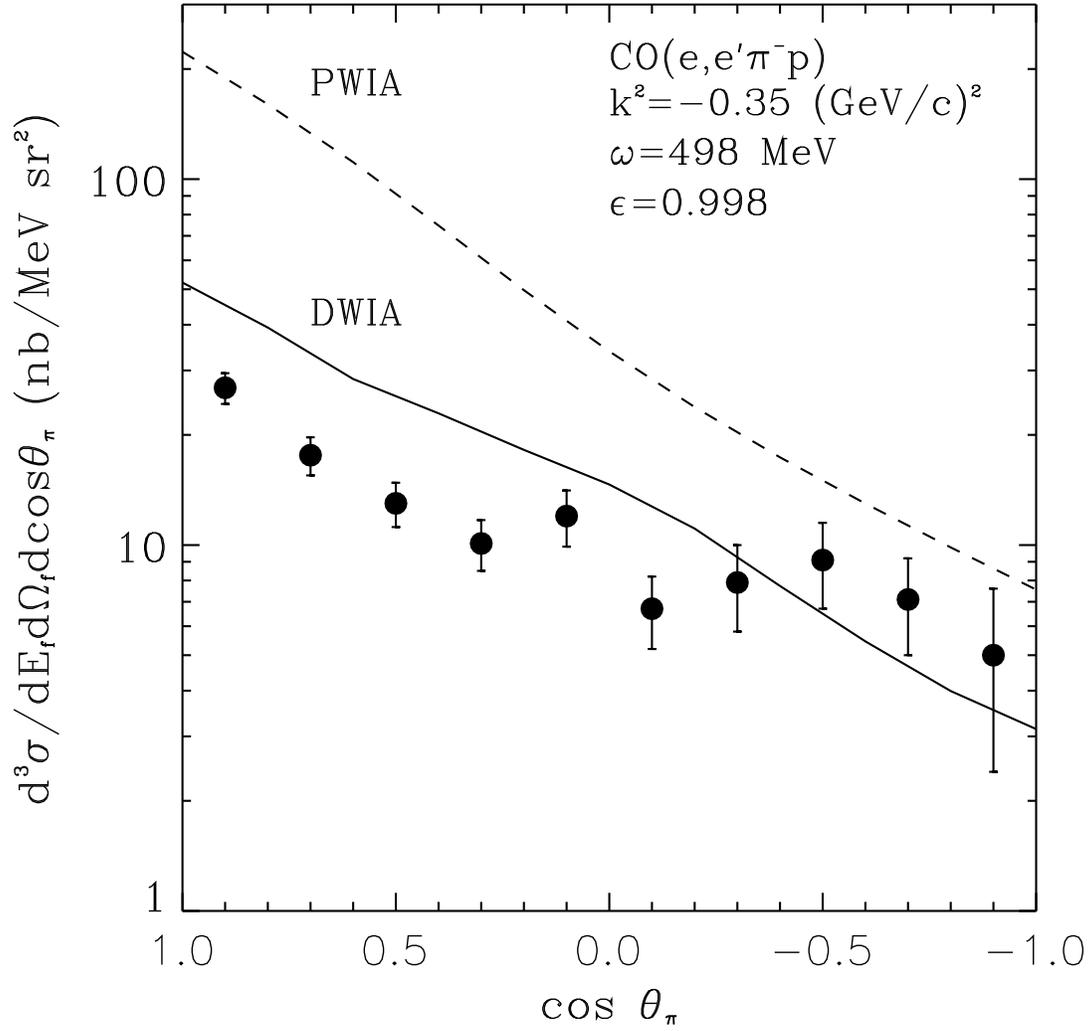,width=14cm}}
\vspace{1cm}
\caption{The $\theta_\pi$ dependence of the calculated cross sections
of the reaction $CO(e,e^\prime \pi^- p)$
are compared with the data from SLAC~\protect\cite{slac}.
The dashed curve is the integrated PWIA result carried out
according to Eq.~(\protect\ref{pwia.slac}). The solid curve is the
DWIA result calculated according to the approximation given by
Eq.~(\protect\ref{dwia.slac}).}
\label{slac}
\end{figure}

Fig.~\ref{slac} depicts the pion angular distribution of the
integrated cross section in the laboratory frame
relative to the direction of the virtual photon momentum for
CO$(e,e^\prime p \pi^-)$ at W=1.210 GeV, and
$-k^2=0.35 \mbox{GeV/c}^2$.
In obtaining the cross sections for the CO target, we have summed over
s-shell and p-shell knockout for $^{12}$C and $^{16}$O.
The PWIA result clearly overestimates the data. The DWIA results
roughly agree with the data at backward angles, but overpredict at
forward pion angles.
Fig.~\ref{slac} also shows that the fall-off of the cross section with
increasing pion angle is more gradual than predicted in either
calculation. At forward angles the data are suppressed relative to DWIA by
approximately a factor of two. This is reminiscent of the results of the
Bates experiment~\cite{bates} for real photons,
where our calculations along with other studies found good
agreement with the data at backward angles and were approximately
three times larger than the data at forward angles.
The difference between forward and backward pion angles may be
due to the structure of the electroproduction operator, which is
predominantly the resonant $\Delta$ term at forward angles and
non-resonant Born terms at backward angles. This would suggest that
the discrepancy at forward angles is due to the $\Delta$-nucleus
interaction. However, full DWIA calculations have to be carried out
before any definite conclusions can be drawn.

\section{Summary}
\label{sumcon}

In this study we have established a DWIA formalism for
calculating  quasifree pion electroproduction
from nuclei, $A(e,e^\prime \pi N)B$, in the $\Delta$ region.
The reaction provides the same promise in studying $\Delta$
excitations in nuclei as does the photoproduction reaction
$A(\gamma, \pi N)B$, with the added advantage and complexity
generated by the longitudinal polarization of the virtual photon.
The sensitivity to the nuclear structure of the target is minimal.
The only information required of the target is the
single particle bound wave-function, the spectroscopic factor,
and the optical potentials.
Kinematically, the reaction provides a great deal of
flexibility since the
target can take up a wide range of momentum transfer
but little recoil energy.

We obtained the pion electroproduction operator by extending the BL
pion photoproduction operator which has enjoyed much success in
pion photoproduction from nuclei. It is gauge invariant,
frame independent,
and simple to implement in a non-relativistic description of nuclei.
We compared the operator to a large body of data on the single
proton and found overall good agreement.

Nuclear cross sections are predicted under a variety of
kinematic situations to help expose the different aspects of
the reaction. Such issues as the interference between the
$\Delta$ resonance and the Born background,
sensitivities to the quadrupole component of the $\Delta$ excitation,
and sensitivities to the electromagnetic form factors, are studied.
Reasonably large sensitivities to the quadrupole component
are found in the neutral pion channel.
Methods on how to separate
the various structure functions in the cross section are suggested.
These results are expected to be useful in planning future experiments.

The first measurements of the reaction in the $\Delta$ region,
a SLAC experiment on $CO(e,e^\prime p \pi^-)$,
are analyzed using our framework.
Preliminary results show that it reveals a similar pion forward-backward
anomaly first found in the Bates experiment
on $^{16}O(\gamma, p \pi^-)^{15}O$.
Clearly, more theoretical and experimental studies are needed at this
point.  The SLAC experiment suffers from low statistics and poor energy
resolution. As a result, the cross sections had to be integrated
over, thus making comparisons with calculations difficult and ambiguous.
However, the situation is expected to improve as
more exclusive data become available in the near future
with machines at MAMI, Bates, LEGS, NIKHEF,
and the commissioning of CEBAF.

\acknowledgements
F.X.L. was supported by the National Sciences and
Engineering Research Council of Canada;
L.E.W. by U.S. DOE under Grant No. DE-FG-02-87-ER40370
and C.B. by U.S. DOE under Grant No. DE-FG02-86-ER40907.
We also acknowledge support from a NATO Collaborative Research Grant
and from the Ohio Supercomputing Center for time on the Cray Y-MP.


\newpage
\appendix
\section{Pion electroproduction operator}
\label{blvirtual}

In this appendix, we give the
full operator for both charged and neutral
pion electroproduction.
The operator is a straightforward extension of the Blomqvist-Laget pion
photoproduction
operator with appropriate form factors and gauge correction terms introduced.
It is given by
 $t_{\gamma_v \pi}=\mbox{\boldmath $\varepsilon_\lambda$}\cdot {\bf J}$
where $\mbox{\boldmath $\varepsilon_\lambda$}$ is the virtual photon
polarization vector and
${\bf J}$ is the pion electroproduction current.
We decompose the operator into spin 0 and spin 1 terms
by writing
\begin{equation}
t_{\gamma_v \pi}(\lambda, {\bf k}, {\bf p}_i ,{\bf q}, {\bf p}) =
L+i\,{\mbox{\boldmath $\sigma$}} \cdot  {\bf K}.
\end{equation}
The non-spin flip term L and the spin flip term ${\bf K}$ each
consists of a coherent sum of the Born and $\Delta$ resonance
terms:
\begin{equation}
L=L_{Born}+L_{\Delta}
\end{equation}
\begin{equation}
{\bf K}={\bf K}_{Born}+{\bf K}_{\Delta}.
\end{equation}
The Born terms in PV coupling for various production channels are
given by the following.
\\
For $\gamma_v$ + p $\rightarrow$ ${\pi}^{+}$ + n:
\begin{equation}
L_{Born}=\frac{\sqrt{2}\,eg_{0}}{2m}\left[ \frac{G^p_M(k^2)}{2E_a
(P^0_a -E_a )} +\frac{G^n_M(k^2)}{2E_b (P^0_b -E_b )} \right]\;
{\bf q} \cdot ({\bf k}\times\mbox{\boldmath $\varepsilon_\lambda$}),
\end{equation}
\begin{eqnarray}
{\bf K}_{Born} & = & \frac{\sqrt{2}\,eg_{0}}{2m}\, \left\{ \left[
-F_A(k^2)+F^p_1(k^2) \,\frac{mE_{\pi}}{E_a (P^0_a +E_a )} \right]\;
\mbox{\boldmath $\varepsilon_\lambda$} \right. \nonumber \\
& & \qquad \left.
+ F_{\pi}(k^2)\,\frac{({\bf k}-{\bf q}\,)\,
\mbox{\boldmath $\varepsilon_\lambda$}\cdot(2{\bf q}-{\bf k})}{(q -k )^2
-m^2_\pi }
- F^p_1(k^2) \,\frac{{\bf q}\;\mbox{\boldmath $\varepsilon_\lambda$}
\cdot (2{\bf p}_i+{\bf k})}{2E_a (P^0_a -E_a )} \right. \nonumber \\
& & \qquad \left.
+ \left[ \frac{G^p_M(k^2)}{2E_a (P^0_a -E_a )}
-\frac{G^n_M(k^2)}{2E_b (P^0_b -E_b )} \right]\;
{\bf q} \times ({\bf k}\times \mbox{\boldmath $\varepsilon_\lambda$}) \right\}.
\end{eqnarray}
\\
For $\gamma_v$ + n $\rightarrow$ ${\pi}^{-}$ + p:
\begin{equation}
L_{Born}=\frac{\sqrt{2}\,eg_{0}}{2m}\left[ \frac{G^n_M(k^2)}{2E_a
(P^0_a -E_a )} +\frac{G^p_M(k^2)}{2E_b (P^0_b -E_b )} \right]\;
{\bf q} \cdot ({\bf k}\times\mbox{\boldmath $\varepsilon_\lambda$}),
\end{equation}
\begin{eqnarray}
{\bf K}_{Born} & = & \frac{\sqrt{2}\,eg_{0}}{2m}\, \left\{ \left[
F_A(k^2) + F^p_1(k^2) \,\frac{mE_{\pi}}{E_b (P^0_b +E_b )} \right]\;
\mbox{\boldmath $\varepsilon_\lambda$} \right. \nonumber \\
& & \qquad \left.
- F_{\pi}(k^2)\,\frac{({\bf k}-{\bf q}\,)\,
\mbox{\boldmath $\varepsilon_\lambda$}\cdot(2{\bf q}-{\bf k})}
{(q -k )^2 -m^2_\pi }
- F^p_1(k^2) \,\frac{{\bf q}\;
\mbox{\boldmath $\varepsilon_\lambda$}
\cdot (2{\bf p}-{\bf k})}{2E_b (P^0_b -E_b )} \right. \nonumber \\
& & \qquad \left.
+ \left[ \frac{G^n_M(k^2)}{2E_a (P^0_a -E_a )}
-\frac{G^p_M(k^2)}{2E_b (P^0_b -E_b )} \right]\;
{\bf q} \times ({\bf k}\times \mbox{\boldmath $\varepsilon_\lambda$}) \right\}.
\end{eqnarray}
\\
For $\gamma_v$ + p $\rightarrow$ ${\pi}^{0}$ + p:
\begin{equation}
L_{Born}=\frac{eg_{0}}{2m}\left[ \frac{G^p_M(k^2)}{2E_a (P^0_a -E_a )}
 +\frac{G^p_M(k^2)}{2E_b (P^0_b -E_b )} \right]
\left({\bf q}-\frac{E_{\pi}}{2m}(2{\bf p}_i-{\bf q})\right)
\cdot({\bf k}\times\mbox{\boldmath $\varepsilon_\lambda$}),
\end{equation}
\begin{eqnarray}
{\bf k}_{Born} & = & \frac{eg_{0}}{2m}\, \left\{
\left[ \frac{G^p_M(k^2)}{2E_a (P^0_a -E_a )}
- \frac{G^p_M(k^2)}{2E_b (P^0_b -E_b )}\right]
\left({\bf q}-\frac{E_{\pi}}{2m}(2{\bf p}_i-{\bf q})\right)
\times({\bf k}\times
\mbox{\boldmath $\varepsilon_\lambda$}) \right. \nonumber \\
& & \left.
- F^p_1(k^2) \,\frac{\mbox{\boldmath $\varepsilon_\lambda$}
\cdot(2{\bf p}_i +{\bf k})}{2E_a (P^0_a -E_a )}
\left[{\bf q}-\frac{E_{\pi}}{2m}({\bf q}+2{\bf p})\right]
\right. \nonumber  \\
& &  \left.
- F^p_1(k^2) \,\frac{\mbox{\boldmath $\varepsilon_\lambda$}
\cdot(2{\bf p}-{\bf k})}{2E_b (P^0_b -E_b )}
\left[{\bf q}-\frac{E_{\pi}}{2m}(2{\bf p}_i-
{\bf q})\right] \right. \nonumber \\
& &  \left.
+ \mbox{\boldmath $\varepsilon_\lambda$}
\left[ \frac{m}{E_a (P^0_a +E_a )}
\left(E_{\pi}-\frac{(2{\bf p}+{\bf q})\cdot{\bf q}}{2m} \right)
\left(F^p_1(k^2) -\frac{E_{\pi}}{2m}\,F^p_2(k^2)\right) \right]
 \right. \nonumber \\
& &  \left. \hspace{-6pt}
+ \mbox{\boldmath $\varepsilon_\lambda$}
\left[ \frac{m}{E_b (P^0_b +E_b )}
\left(E_{\pi}-\frac{(2{\bf p}_i -{\bf q})\cdot{\bf q}}{2m} \right)
\left(F^p_1(k^2) +\frac{E_{\pi}}{2m}\,F^p_2(k^2)\right)\right] \right\}.
\end{eqnarray}
\\
For $\gamma_v$ + n $\rightarrow$ ${\pi}^{0}$ + n:
\begin{equation}
L_{Born}=\frac{eg_{0}}{2m}\left[ \frac{G^n_M(k^2)}{2E_a (P^0_a -E_a )}
 +\frac{G^n_M(k^2)}{2E_b (P^0_b -E_b )} \right]
\left({\bf q}-\frac{E_{\pi}}{2m}(2{\bf p}_i-{\bf q})\right)
\cdot({\bf k}\times\mbox{\boldmath $\varepsilon_\lambda$}),
\end{equation}
\begin{eqnarray}
{\bf K}_{Born} &=& \frac{eg_{0}}{2m}\, \left\{
\left[\frac{G^n_M(k^2)}{2E_a (P^0_a -E_a )}
- \frac{G^n_M(k^2)}{2E_b (P^0_b -E_b )} \right]
\left({\bf q}-\frac{E_{\pi}}{2m}(2{\bf p}_i-{\bf q})\right)
\times({\bf k}\times\mbox{\boldmath $\varepsilon_\lambda$})
 \right. \nonumber \\
& &  \left.
+ \mbox{\boldmath $\varepsilon_\lambda$}
\left[ \frac{m}{E_a (P^0_a +E_a )}
\left(E_{\pi}-\frac{(2{\bf p}+{\bf q})\cdot{\bf q}}{2m} \right)
\left(-\frac{E_{\pi}}{2m}\,F^n_2(k^2) \right) \right]
\right. \nonumber \\
& &  \left.
+ \mbox{\boldmath $\varepsilon_\lambda$}
 \left[ \frac{m}{E_b (P^0_b +E_b )}
\left(E_{\pi}-\frac{(2{\bf p}_i -{\bf q})\cdot{\bf q}}{2m} \right)
\left(\frac{E_{\pi}}{2m}\, F^n_2(k^2) \right) \right] \right\}.
\end{eqnarray}
The photon, incoming nucleon,
pion and outgoing nucleon four-momenta are
$k=(\omega,{\bf k}\,)$, $p_i=(E_{p_i}, {\bf p}_i\,)$,
$q=(E_{\pi}, {\bf q}\,)$,
$p=(E_N, {\bf p}\,)$, respectively. They can be
in any reference frame.
The nucleon mass is denoted as m.
 The four-momenta in the s and u
channels are $P_a =k+p_i $
and $P_b = p_i -q=p-k$
and $E_{a,b}=(|{\bf p}_{a,b}|^2 + m^2 )^{1/2}$.
For the $\pi-N$ coupling constant we use $g^2_0 /4\pi =14$.

In the $\pi^0$ channels, the following
$\omega$-exchange term should be added coherently to L.
\begin{equation}
L_\omega = \frac{1}{m_\pi}\;
\frac{eg_{\omega_1}\,F_{\omega}(k^2)}
{(q^\mu -k^\mu )^2 -m^2_\omega}\;
{\bf q} \cdot ({\bf k}\times\mbox{\boldmath $\varepsilon_\lambda$})
\end{equation}
where $m_\omega=750$ MeV,
$g_{\omega_1}=10$.

The $\Delta$ resonance terms with M1 and E2 and C2 transitions
are given by
\begin{equation}
L_{\Delta}=
\frac{eC_{\pi}C_{\gamma}G_1 G_3\,F_{\Delta}(k^2)\,e^{i\phi_M}}
{P^2_a -M_{\Delta}^2 +i\,\Gamma_\Delta \,M_{\Delta}}
\times \frac{2}{3} {\bf q}^{\ast}\cdot ({\bf k}^{\ast}\times
\mbox{\boldmath $\varepsilon_\lambda$} ).
\end{equation}
\begin{eqnarray}
{\bf K}_{\Delta} & = &
\frac{-eC_{\pi}C_{\gamma}G_1 G_3\,F_{\Delta}(k^2)}
{P^2_a -M_{\Delta}^2 +i\,\Gamma_\Delta \, M_{\Delta}} \left\{ \,
\frac{1}{3}{\bf q}^{\ast}\times ({\bf k}^{\ast}\times
\mbox{\boldmath $\varepsilon_\lambda$})\,
e^{i\phi_M} \right. \nonumber \\
& & \left. +
\left[\, ({\bf q}^{\ast}\cdot {\bf k}^{\ast})\,
\mbox{\boldmath $\varepsilon_\lambda$}
+ ({\bf q}^{\ast}\cdot \mbox{\boldmath $\varepsilon_\lambda$})\,
{\bf k}^{\ast} + \frac{2}{3}\,\frac{M_{\Delta}-m}{m}\,
(\mbox{\boldmath $\varepsilon_\lambda$} \cdot {\bf p}_i)\;
 {\bf q}^{\ast} \right. \right. \nonumber \\
& & \left. \left.
-\frac{2\mbox{\boldmath $\varepsilon_\lambda$} \cdot {\bf k}}
{|{\bf k}|^2}\,({\bf q}^{\ast}\cdot {\bf k}^{\ast})\, {\bf k}^{\ast}
\right]\, \frac{2\omega m \alpha}{(3M_{\Delta}+m)(M_{\Delta}+m)}\,
e^{i\phi_E } \right. \nonumber \\
& & \left. \hspace{-10pt}
+ \frac{2\mbox{\boldmath $\varepsilon_\lambda$} \cdot {\bf k}}
{3|{\bf k}|^2}\,\left[3({\bf q}^{\ast}\cdot {\bf k}^{\ast})\,
 {\bf k}^{\ast}-|{\bf k}|^2\,{\bf q}^{\ast}\,\right]\,
\frac{2\omega m \alpha}{(3M_{\Delta}+m)(M_{\Delta}+m)}\,
\frac{k^{\gamma}}{|{\bf k}^{\ast}|} \right\} \label{alpha1}
\end{eqnarray}
where ${\bf k}^{\ast}={\bf k}-(M_{\Delta}-m){\bf p}_i /m$ and
${\bf q}^{\ast}={\bf q}-E_\pi ({\bf k}+{\bf p}_i) /M_{\Delta}$,
the isospin coefficients
$C_{\pi^{\pm}}C_{\gamma}=\mp \frac{\sqrt{2}}{3}$ and
$C_{\pi^{0}}C_{\gamma}=\frac{2}{3}$.
The coupling constants $G_1, G_3$, the mass of the delta
$M_{\Delta}$ and the width of the delta $\Gamma_\Delta$
were treated as parameters and fitted to the data.
We use the following parameterization~\cite{bl1}
\begin{eqnarray}
M_\Delta &=& 1225 \;\mbox{MeV}, \nonumber \\
\Gamma_\Delta &=& 110 {\left( \frac{|{\bf q}|}{|{\bf p}_a|}
\right)}^3
\frac{M_\Delta}{P_a}
\frac{1+(0.007|{\bf p}_a| )^2}{1+(0.007|{\bf q}|)^2} \;
\mbox{MeV},\nonumber \\
G_1 &=& 0.34\frac{M_\Delta +m}{m_\pi}, \nonumber \\
G_3 &=& 2.18/{m_\pi}\; {\mbox{MeV}}^{-1}
\end{eqnarray}
where ${\bf p}_a$ and ${\bf q}$ are in MeV/c and $m_\pi$ is in MeV.
The phases used to restore unitarity are functions
of the variable $x=P_a -1080$ in MeV and
are given in degrees as
\begin{eqnarray}
\phi_M &=& -0.1228\sqrt{x}+ 0.0735x, \nonumber \\
\phi_E &=& 3.9136\sqrt{x}+0.2795x- 0.00049x^2.
\end{eqnarray}
The symbol $\alpha$ in Eq.~(\ref{alpha1})
is a constant that measures the relative strength between
the M1 and E2 transition amplitudes in the $\Delta$
 and here takes the value
$\alpha=0.8$, while $k^{\gamma}=(P_a^2-m^2)/(2P_a)$
 is the equivalent photon energy in the $\pi$-N c.m. frame.

In the following we give the gauge corrections for the Born terms.
They should be added to the Born part of the
$\mbox{\boldmath $\varepsilon_\lambda$}\cdot {\bf J}$ for various
channels. The $\Delta$ term is made separately gauge invariant and no
additional terms need to be added. \\
For $\gamma_v$ + p $\rightarrow$ ${\pi}^{+}$ + n:
\begin{eqnarray}
\mbox{\boldmath $\varepsilon_\lambda$}\cdot {\bf J}^{(G)}
& = &
\frac{\mbox{\boldmath $\varepsilon_\lambda$}\cdot {\bf k}}{-k^2}\,
 \frac{\sqrt{2}\,eg_{0}}{2m}\;i\,\mbox{\boldmath $\sigma$} \cdot
\left\{ \left[ (F_\pi -F^p_1 )-\frac{\omega}{2m}(F^p_1 -F_A )
\right]\, {\bf q} + \frac{\omega}{m}(F^p_1 -F_A )\,{\bf p}_i
\right. \nonumber \\
& & \left. \hspace{-2.5cm}
+ \left[ (F_A -F_\pi )+\frac{\omega}{2m}(F^p_1 -F_A)
+F^p_1 \left(\frac{\omega E_\pi}{2E_a (P^0_a -E_a )}
-\frac{mE_\pi}{E_a (P^0_a +E_a )} \right) \right]{\bf k} \right\}.
\end{eqnarray}
\\
For $\gamma_v$ + n $\rightarrow$ ${\pi}^{-}$ + p:
\begin{eqnarray}
\mbox{\boldmath $\varepsilon_\lambda$}\cdot {\bf J}^{(G)}
\hspace{-2.5pt} & = &  \hspace{-2.5pt}
\frac{\mbox{\boldmath $\varepsilon_\lambda$}\cdot {\bf k}}{-k^2}\,
 \frac{\sqrt{2}\,eg_{0}}{2m}\;i\,\mbox{\boldmath $\sigma$} \cdot
\left\{ -\left[ (F_\pi -F^p_1 )-\frac{\omega}{2m}(F^p_1 -F_A )
\right]\, {\bf q} - \frac{\omega}{m}(F^p_1 -F_A )\,{\bf p}_i
\right. \nonumber \\
& & \left. \hspace{-2.4cm}
+ \left[-(F_A -F_\pi )-\frac{\omega}{2m}(F^p_1 -F_A)
+F^p_1 \left(\frac{\omega E_\pi}{2E_b (P^0_b -E_b )}
+\frac{mE_\pi}{E_b (P^0_b +E_b )} \right) \right]{\bf k} \right\}.
\end{eqnarray}
\\
For $\gamma_v$ + p $\rightarrow$ ${\pi}^{0}$ + p:
\begin{eqnarray}
\mbox{\boldmath $\varepsilon_\lambda$}\cdot {\bf J}^{(G)} & = &
\frac{\mbox{\boldmath $\varepsilon_\lambda$}\cdot {\bf k}}{-k^2}\,
 \frac{eg_{0}}{2m}\;i\,\mbox{\boldmath $\sigma$} \cdot \left\{
\left[-F^p_1 \left( \frac{\omega E_\pi}{2E_a (P^0_a -E_a )}
+ \frac{\omega E_\pi}{2E_b (P^0_b -E_b )} \right)
\right. \right. \nonumber \\
& & \left. \left. \hspace{-55pt}
-F^p_1 \frac{2m\omega -(|{\bf k}|^2+2{\bf p}_i \cdot{\bf k})
(1+\frac{E_\pi}{2m})}{2E_a (P^0_a -E_a )}
+F^p_1 \frac{-2m\omega +(-|{\bf k}|^2+2{\bf p} \cdot{\bf k})
(1+\frac{E_\pi}{2m})}{2E_b (P^0_b -E_b )} \right]{\bf q}
\right. \nonumber \\
& &  \left.
+ \left[F^p_1\frac{E_\pi}{m}
\frac{2m\omega -|{\bf k}|^2-2{\bf p}_i \cdot{\bf k}}
{2E_a (P^0_a -E_a )}
- F^p_1\frac{E_\pi}{m}
\frac{-2m\omega -|{\bf k}|^2+2{\bf p} \cdot{\bf k}}
{2E_b (P^0_b -E_b )} \right]{\bf p}_i \right. \nonumber \\
& & \left. \hspace{-3pt}
+ \left[F^p_1\frac{E_\pi}{m}
\frac{2m\omega -|{\bf k}|^2-2{\bf p}_i \cdot{\bf k}}
{2E_a (P^0_a -E_a )}
- F^p_1\left(\frac{\omega E_\pi}{2E_a (P^0_a +E_a )}
- \frac{\omega E_\pi}{2E_b (P^0_b +E_b )} \right)
\right. \right. \nonumber \\
& & \left. \left.
- \frac{m}{E_a (P^0_a +E_a )}\left(E_\pi -\frac{(2{\bf p}
+{\bf q}) \cdot {\bf q}}{2m} \right)\left(F^p_1 -\frac{E_\pi}{2m}F^p_2 \right)
\right. \right. \nonumber \\
& & \left. \left.
- \frac{m}{E_b (P^0_b +E_b )}\left(E_\pi -\frac{(2{\bf p}_i
-{\bf q}) \cdot {\bf q}}{2m} \right)\left(F^p_1 +\frac{E_\pi}{2m}F^p_2 \right)
\right]{\bf k} \right\}.
\end{eqnarray}
\\
For $\gamma_v$ + n $\rightarrow$ ${\pi}^{0}$ + n:
\begin{eqnarray}
\mbox{\boldmath $\varepsilon_\lambda$}\cdot {\bf J}^{(G)} & = &
\frac{\mbox{\boldmath $\varepsilon_\lambda$}\cdot {\bf k}}{-k^2}\,
 \frac{eg_{0}}{2m}\;i\,\mbox{\boldmath $\sigma$} \cdot \left\{
F^p_2 \left[\frac{E_\pi}{2E_a (P^0_a +E_a )}\left(E_\pi -\frac{(2{\bf p}
+{\bf q}) \cdot {\bf q}}{2m} \right) \right. \right. \nonumber \\
& & \left. \left.
- \frac{E_\pi}{2E_b (P^0_b +E_b )}\left(E_\pi -\frac{(2{\bf p}_i
-{\bf q}) \cdot {\bf q}}{2m} \right) \right]{\bf k} \right\}.
\end{eqnarray}
Note that these additional terms go to zero (to order $(p/m)^2$)
if the same form factors are used:
$F_A=F^p_1=F_\pi$. The following relations can help demonstrate it:
\begin{eqnarray}
2E_a (P^0_a -E_a ) & \approx & 2m\omega -|{\bf k}|^2 -2{\bf p}_i\cdot {\bf k},
\nonumber \\
2E_b (P^0_b -E_b ) & \approx & -2m\omega -|{\bf k}|^2 +2{\bf p}\cdot {\bf k},
\nonumber \\
2E_a (P^0_a +E_a ) & \approx & 2m(2m+\omega),
\nonumber \\
2E_b (P^0_b +E_b ) & \approx & 2m(2m-\omega).
\end{eqnarray}


\section{Electromagnetic form factors used in the model}
\label{formfactors}

For the nucleon form factors we use the well-known dipole form:
\begin{equation}
\frac{G^p_M (k^2)}{\mu_p}=\frac{G^n_M (k^2)}{\mu_n}=G^p_E (k^2)
=-\frac{G^n_E (k^2)}{\tau\,\mu_n}
=\left(1+\frac{-k^2}{0.71} \right)^{-2},
\end{equation}
where $\mu_p=2.79$, $\mu_n=-1.91$, $\tau=-k^2/(4m^2)$
and $k^2$ is in (GeV/c)$^2$. The form factors $F_1$ and $F_2$
are given in terms of $G_E$ and $G_M$ by
\begin{equation}
F^{p,n}_1 (k^2)=\frac{G^{p,n}_E + \tau\, G^{p,n}_M}{1+\tau},
\end{equation}
\begin{equation}
F^{p,n}_2 (k^2)=\frac{G^{p,n}_M - G^{p,n}_E}{1+\tau}.
\end{equation}
Note that in this definition $F^n_1 (k^2)=0$.
For the pion form factor we use the monopole form
\begin{equation}
F_\pi (k^2)=\left(1+\frac{-k^2}{0.45}\right)^{-1}.
\end{equation}
For the axial form factor we use
\begin{equation}
F_A (k^2)=\left(1+\frac{-k^2}{0.90} \right)^{-2}.
\end{equation}
For the $\Delta$ form factor we use the following form
\begin{equation}
F_\Delta (k^2)=\left(1+\frac{k^2}{6} \right)\,
\left(1+\frac{-k^2}{0.72} \right)^{-2}.
\end{equation}
Finally, the form factor at the $\omega{\pi}^0 \gamma$ vertex is
taken to be of $\rho$(775) type:
\begin{equation}
 F_{\omega}(k^2)=0.374\,\left(1+\frac{-k^2}{0.60} \right)^{-1}.
\end{equation}

The RMS charge radius $\sqrt{<r^2>}$
is related to the form factor by (assume $k^2<0$):
\begin{equation}
<r^2>=-6\,\left. \frac{d\,F(k^2)}{d\,(-k^2)}\right\vert_{\,k^2=0}.
\end{equation}

\newpage

\end{document}